\documentclass[a4paper,fleqn,usenatbib]{mnras}

\usepackage{newtxtext,newtxmath}
\usepackage[T1]{fontenc}
\usepackage{ae,aecompl}

\usepackage{graphicx}	% Including figure files
\usepackage{amssymb}	% Extra maths symbols

\title[Star-formation rates of cluster galaxies: nature $vs$ nurture]{Star-formation rates of cluster galaxies: nature $vs$ nurture}

\author[Lagan\'{a} \& Ulmer]{
Tatiana F. Lagan\'{a},$^{1}$\thanks{E-mail: tatiana.lagana@cruzeirodosul.edu.br}
M. P. Ulmer $^{2}$
\\
$^{1}$NAT, Universidade Cruzeiro do Sul, Rua Galv\~{a}o Bueno, 868, CEP:01506-000, S\ ~{a}o Paulo-SP, Brazil\\
$^{2}$Department of Physics and Astronomy and CIERA, Northwestern University, 2145 Sheridan Road, Evanston, IL 60208-3112, USA\\}
\date{Accepted 2017 December 6. Received 2017 December 1; in original form 2017 July 7}

\pubyear{2015}

\begin{document}
\label{firstpage}
\pagerange{\pageref{firstpage}--\pageref{lastpage}}
\maketitle

\begin{abstract}
We analyzed 17 galaxy clusters, and investigated, for the first time, the dependence of the SFR and sSFR as a function of projected distance 
(as a proxy for environment) and stellar mass for cluster galaxies in an intermediate-to-high redshift range ($0.4 < z < 0.9$). 
We used up to nine flux points (BVRIZYJHKs magnitudes), its errors and redshifts to compute M$_{\rm{star}}$, SFR and sSFR through spectral energy 
distribution fitting technique. We use a z-dependent sSFR value to distinguish star-forming (SF) from quiescent galaxies.
To analyse the SFR and sSFR history we split our sample in two redshift bins: galaxies at $0.4 < z < 0.6$ and $0.6 < z < 0.9$. We separate the effects of environment 
and stellar mass on galaxies by comparing the properties of star-forming and quiescent galaxies at fixed environment (projected radius) and fixed stellar mass.
For the selected spectroscopic sample of more than 500 galaxies, the well-known correlation between SFR and $M_{\rm star}$ is already in place at 
$z \sim 0.9$, for both SF and quenched galaxies. 
Our results are consistent with no evidence that SFR (or sSFR) depends on environment, suggesting that for cluster galaxies at an intermediate-to-high 
redshift range, mass is the primary characteristic that drives SFR.
%Due to mass segregation, high mass systems lie generally in dense regions, while low mass systems prefer low density regions. The high mass systems
%dominate the low sSFR end and these facts lead to a global sSFR-density anti-correlation that is observed in this work.

\end{abstract}


%%%%%%%%%%%%%%%%%%%%%%%%%%%%%%%%%%%%%%%%%%%%%%%%%%

%%%%%%%%%%%%%%%%% BODY OF PAPER %%%%%%%%%%%%%%%%%%

\section{Introduction}
Galaxies are the building blocks of our Universe and to understand their evolution through the history of the Universe is of fundamental importance.
The hierarchical Lambda  Cold Dark Matter ($\Lambda$CDM) model predicts that galaxies are formed through the gravitational collapse of baryons 
(mostly in the form of gas) 
inside dark matter haloes. 

Interestingly, many of the galaxy properties, such as star formation rates (SFRs), morphology and colour show environmental and redshift 
dependence as well. Thus, from an observational point of view, measuring these properties
at different epochs and environments will give us important clues on the evolution of galaxies.
%Locally, galaxies show distinct SF properties  and colours in different environments 
%\citep{Lewis02,Kauffmann04,Blanton05}. 

In 1980, \citet{BO84} made a pioneering study in clusters of galaxies showing that the blue fraction of galaxies increased from $z=0$
to $z \sim 0.5$. \citet{Dressler99} showed the effect needed refinement by defining E$+$A galaxies as `blue'.
More recently, however, a different approach has been used to define galaxy evolution by measuring the star formation rate (SFR) 
or specific star formation rate (i.e., star formation per unit mass, $\rm sSFR = \rm SFR/ \rm M_{star}$).

An important fact is that galaxies in the local universe have lower levels of star-formation (SF) activity than the ones in the past. The global star formation history shows a peak at $z \sim 1$
and then drops towards $z \sim 0$ \citep[][]{Lilly96,Madau96,Karim11,Sobral13,Khostovan15}. Also the SFRs of normal star-forming (SF) galaxies at $z \sim 0$ are significantly lower than the 
SFRs of higher redshift  SF galaxies with similar mass \citep{Daddi07, Sobral14}. Therefore we need to understand the physical mechanism(s) that can lower and eventually 
stop the SF activity of galaxies to explain this change on average  SF properties of galaxies from high redshift to the local universe.

Both internal and environmental factors seem to affect SF activity of galaxies \citep{Peng10,Sobral11,Muzzin12,Darvish15,Darvish16}. The depletion of the gas reservoir, 
and the decline in the galaxy merger rate
have been considered as possible alternatives to explain the reduction in SF activity since $z \sim 1$ \citep{LeFevre00, Hammer05,Noeske07}. But also,
since detailed observations of dense environments such as galaxy groups and clusters in the local universe have shown that galaxies that reside in these environments 
have properties very different from galaxies in low density or field galalaxies,
many environmental effects may be linked to the decrease in the global SF. Processes such as ram-pressure stripping or galaxy harassment, which are dominant
in regions of higher density \citep[e.g.,][]{GG72,Moore96,Hester06} can remove gas from infalling galaxies as they merge into groups and clusters of galaxies, 
leading to a decrease in SF via starvation. Heating of intracluster gas due to cluster mergers \citep[][]{McCarthy07} or viral shock heating of infalling gas
in massive dark matter haloes could also be responsible for the exhaustion of the cold gas supply of galaxies that are in high-density environments. 
\citet{ZM98} have proposed that these particular properties of cluster and group galaxies may be due to `pre-processing' before the accretion into the dense
environment \citep[but see][who claimed that `pre-processing' is not a large effect]{Berrier09}. This is supported by observations of reduced star formation rates of galaxies that 
reside beyond the viral radius \citep{Balogh99,Lewis02}.

From the observational point of view, for field galaxies, there seems to be a consensus in 
that there exists 
a dependence of the SFR on the stellar mass, with more massive galaxies having higher SFRs \citep[e.g.,][]{LaraLopez10, Peng10}. 
A different result was reported by  \citet{Noeske07} who analysed field galaxies and showed that in the SFR-$M_{\rm star}$ relation there are two distinct populations: a main sequence of SF galaxies 
in which SFR depends on $M_{\rm star}$,  and ``quenched'' galaxies  with no detectable SF activity, forming almost an horizontal line in this diagram.

If most of the results agrees for the SFR dependence on $M_{\rm star}$, there has been a major divergence on the environmental dependence 
of SF activity of galaxies.
While several investigations \citep{Gomez03,Balogh04,Kauffmann04} find the mean SFR of galaxies in dense environment to be much 
less than those of galaxies in lower density regions, some authors reported that an inversion of the local
relation is seen at high redshifts \citep[$z \leq 1$][]{Cooper08,Elbaz07}, such that field galaxies in high-density environments have enhanced SFRs.
Part of this discrepancy may be connected with a possible mass dependence: if mass-downsizing is already in place at higher redshifts \citep{Popesso11,Sobral11,Scodeggio09}, 
and massive galaxies are preferentially located in high-density regions, samples with different luminosity limits or that have different selection may find contradictory results.
%\citet{Peng10} was the first to disentangle the mass-driven and the environment driven evolution for galaxies up to $z \sim 1$. These authors could distinguish two processes that
%affect the  galaxy  star  formation  activity: ``mass-quenching'' and ``environment-quenching''.

Extending the field work to clusters of galaxies at higher redshifts than $z \sim 0.4-0.5$ has required assembling significant amount of data of relatively
faint galaxies. Therefore only in the past few years have such data sets become available. 
For cluster galaxies, the dependence of the SFR on $M_{\rm star}$ seems to still persist but the SFR-density for a single cluster was found to be very weak 
\citep[e.g.,][]{PC13,Lagana16}. Although to determine if these results, for a relatively young cluster, is representative of high-redshift systems, more investigation
is needed.

Thus, we have taken advantage of the existence of one such data set of 
17 rich clusters of galaxies \citep{Guennou10,Martinet15} to carry out the study of the SF history in cluster galaxies.
%Therefore more investigation using a sample of cluster galaxies 
%up to high redshift is needed. 
%We selected a sample of 17 DAFT/FADA clusters to analyse the SFR and sSFR dependence on stellar mass and environment.
Our aim is to determine the form of sSFR/SFR-$M_{\rm star}$ and sSFR/SFR-density relation for the first sample of intermediate-to-high redshift of cluster galaxies
and compare it to the results discussed above. Since our sample covers a wide range in redshift we could separate them into two 
redshift bins: $0.4 < z < 0.6$ and $0.6 < z < 0.9$ to analyse a possible evolution of these relations.
 
The paper is organised as follows. In Section \ref{sample} we describe our sample and describe how $M_{\rm star}$, SFR and sSFR were computed.
 We present our results and in Section \ref{res}, the discussion in Section \ref{disc}, and present our conclusions in Section \ref{conc}.
Throughout this paper we assume concordant $\Lambda$CDM model with $\Omega_{m}$ =0.3, $\Omega_{\Lambda}=0.7$, 
$H_{0} =71 \; \rm km \; s^{-1} \; Mpc^{-1}$.

\section{The data sample and analysis}
\label{sample}
We selected 17 galaxy clusters from DAFT/FADA survey \citep{Guennou10,Martinet15}, presented in Tab. \ref{tab:sample}
in the intermediate-to-high-redshift ($0.4 < z < 0.9$). DAFT/FADA is based on a sample of 91 
high redshift ($z > 0.4$), massive ($>3 \times 10^{14} M_{\odot}$) 
clusters with existing HST imaging, for which  complementary multi-wavelength imaging were performed. 

We selected only galaxies spectroscopically confirmed as cluster members.
The average galaxy 90\% completeness limit of our sample in the I-band is 23.2,
\citep[see][for more information]{Martinet15}, and we used the method presented in \citet{Lagana13}
and described in  Appendix \ref{Mstarz} for 
the adopted stellar mass limit we relied on. We thus considered only galaxies  with $\log(M_{\rm star}/M_{\odot}) > 9.4$

\begin{table*}
\scriptsize
	\centering
	\caption{Cluster galaxies analysed in this work. The different columns correspond to\#1: cluster ID; \#2: right ascension; \#3: declination; \#4: redshift; 
	\#5: scale (kpc/arcsec);\#6: number of galaxies belonging to the cluster after completeness limit; \#7: total masses in $r_{500}$, $M_{500}^{X}$, 
	derived from XMM X-ray data from \citet{Lagana13} or \citet{Guennou14} when
	denoted by symbol $^{\rm G14}$; \#8: total masses in $r_{500}$, $M_{500}^{X}$, homogenised by \citet{Piffaretti11}; \#9:  total masses in $r_{500}$, $M_{500}^{WL}$, 
	derived from weak lensing analysis by \citet{Sereno15};  \#10: total masses in $r_{200}$, $M_{200}^{WL}$, derived fromfrom weak lensing analysis by \citet{Sereno15}.}
	\label{tab:sample}
	\begin{tabular}{cccccccccc} % four columns, alignment for each
		\hline
Cluster & R.A. & DEC & z & scale & $N_{z}$ & $M_{500}^{X}$ & $M_{500}^{X}$ &  $M_{500}^{WL} $ &  $M_{200}^{WL} $\\
& & & &kpc/arcsec & & ($10^{14} h_{70}^{-1} M_{\odot}$) & ($10^{14} h_{70}^{-1} M_{\odot}$) & ($10^{14} h_{70}^{-1} M_{\odot}$)  &  ($10^{14} h_{70}^{-1} M_{\odot}$) \\
\hline
CL 0016$+$1609				&	00 18 33.3	&	$+$16 26 35.8	&	0.546	&6.411 & 64  & - 			   &  & 3.929 & 6.067 \\
CL J0152.7$-$1357				&	01 52 41.0	&	-13 57 45.0	&	0.831	&7.603 & 67   & 1.77 $\pm$ 0.40 & - &2.577 & 3.683 \\
ABELL 0851					&	09 42 56.6	&	$+$46 59 21.9	&	0.407	&5.430 & 144 & 5.5 $\pm$ 1.2$^{\rm G14}$ & - & 6.100 & 9.000 \\
LCDCS 0130					&	10 40 41.6	&	-11 55 51.0	&	0.704	&7.146 & 15 	  & - 			    &-  & 0.405 & 0.640 \\
LCDCS 0173					&	10 54 43.5	&	-12 45 50.0	&	0.750	&7.338 & 14    & - 			    & - & 2.937 & 4.643 \\
MS 1054-03					&	10 57 00.2	&	-03 37 27.4	&	0.823	&7.580 & 55    & - 			      &  3.39 & 6.941 & 11.100\\
RXC J1206.2-0848				&	12 06 12.0	&	-08 48 00.0	&	0.440	&5.533 & 28    & -  		      &  8.40 & 11.200 & 16.100\\
LCDCS 0504					&	12 16 45.1	&	-12 01 17.0	&	0.794	&7.476 & 6     & -  		      & - & 5.835 & 9.226 \\
BMW-HRI J122657.3$+$333253	&	12 26 58.0	&	$+$33 32 54.1	&	0.890	&7.765 & 13    & 12.1 $\pm$ 0.4$^{\rm G14}$ & - & - & - \\
LCDCS 0531					&	12 27 53.9	&	-11 38 20.0	&	0.636	&6.861 & 27    & - 			       & - & 0.310 & 0.491 \\
HDF:ClG J1236$+$6215			&	12 37 60.0	&	$+$62 15 54.0	&	0.850	&7.658 & 15    & - 			       & - & 1.662 & 2.628 \\
LCDCS 0829					&	13 47 32.0	&	-11 45 42.0	&	0.451	&6.411 & 21    & 16.9 $\pm$ 3.6$^{\rm G14}$ & - & 13.557 & 13.886 \\
LCDCS 0853					&	13 54 09.5	&	-12 30 59.0	&	0.763	&7.373 & 9    & - 			       &-  & - & - \\
3C 295 CLUSTER				&	14 11 20.2	&	$+$52 12 09.0		&	0.460	&5.831 & 60    & - 			       & - & 7.200 & 10.527 \\
MACS J1423.8$+$2404			&	14 23 48.3	&	$+$24 04 47.0	&	0.545	&6.382 & 8     & 4.30 $\pm$ 0.72    &  4.98 & 5.826 & 8.107 \\
MACS J1621.4$+$3810			&	16 21 24.0	&	$+$38 10 02.0	&	0.465	&5.831 & 15    & 4.28 $\pm$ 0.35    &  4.81 & 7.044 & 10.181\\
MACS J2129.4$-$0741			&	21 29 26.0	&	-07 41 27.0	&	0.589	&6.628 & 5       & 6.06  	$\pm$ 0.80 &  7.66 &13.486 & 19.490 \\		
\hline
total number of galaxies & & & & & 536 & & \\
\hline
\end{tabular}
\end{table*}

For the selected clusters,  we used a data set with up to nine flux points (BVRIZYJH$K_{s}$ magnitudes), 
its errors and the redshift to compute stellar population synthesis models. 
We stress that we use more than one near infrared band (at least Z and J or Ks) to constrain the IR regime. 
We did not use far infra-red (FIR) because Spitzer data are unavailable for the entire sample. Another problem with the Spitzer data '
is the small angular extent of our clusters, 
for which typical galaxy-galaxy separations are often smaller than the IRAC spatial resolution, leading to considerable 
confusion in the central parts of clusters (see Appendix \ref{ApendixSED} for discussion).

%There are IRAC1 and IRAC2 data available for five clusters and the SED fits with these bands and without IRAC bands are consistent.
%We discuss them in Appendix \ref{ApendixSED}. Also, in  Appendix \ref{ApendixSED}, we show
%SED fits for high and low redshift galaxies considered as star-forming and quenched.}

$\rm M_{star}$, SFR and the sSFR
 were obtained using spectral energy distribution fitting technique performed by MAGPHYS \citep{daCunha08}. 
It is based on the stellar population synthesis 
 models of \citet{BC03}, with a \citet{Chabrier03} 
stellar initial mass function and a metallicity value in the range 0.02-2 $Z_{\odot}$. We set all member galaxies to the mean cluster redshift and for
each galaxy model MAGPHYS produces both the dust free (unattenuated) and attenuated spectrum. The attenuated spectra is obtained using the
dust model of \citet{CF00}. 

 For each galaxy the observed flux points are compared to model flux points, and the goodness of the fit ($\chi^{2}$) determines the probability weight for the given model,
and thus of the associated model parameters in the final probability distribution of each parameter. We adopted the best-fit value as our fiducial estimate
of a given parameter, with lower and upper limits provided by 16\% and 84\% percentiles. Using these limits we find that a typical 1$\sigma$ error is about
10\% and we used this value as representative of our errors. 
In  Appendix \ref{ApendixSED}, we show
SED fits for high and low redshift galaxies considered as star-forming and quenched.
%median value of the probability distribution as our fiducial estimate.

To analyse our correlations, we applied a robust  (that is an outlier-resistant) bootstrap linear fit to data with 10,000 resampling to compute the bisector of the Y $vs$ X
 and X $vs$ Y regression and a Pearson correlation coefficient.  The presented linear fits in all figures are the mean linear fit of the 10,000 fits performed.

\section{Results}
\label{res}

To study if the global trends observed locally were in place at $z \sim 0.9$, we show  the SFR/sSFR against projected radial distance
and against stellar mass. But first, we are interested in separating star-forming (SF) from non-star forming (quiescent) galaxies. 

Since morphological type is related with star-formation, 
colour index  has been commonly used as an early  and late type morphological segregator \citep{Faber07,Lee07} that would correspond
to distinguish quiescent from star-forming galaxies. 

However, there are some caveats related to this method. 
First, depending on the selected bands, there might be up to 30\% of misclassification \citep[e.g. using u-r as reported in][]{Strateva01}.
Also \citet{Wolf05} have described a population of  `dusty red galaxies' dominating the outskirts of the cluster A901/A902. Their sample has red colours due to dust 
and intermediate age rather than old age as for regular early types. As reported by  \citet{Lee15}, at the redshift range of the galaxies in this work,
 there might be a non-negligible red fraction that are still forming stars
and would be misclassified according to a colour selection. 
These red SF galaxies are, on average, dustier, and are migrating into the red quiescent population. However, according to these authors, 
no red SF galaxy has a sSFR $>10^{-8} \; \rm yr^{-1}$.

Having said that, our approach to separate SF from quiescent galaxies is to use the sSFR. 
The sSFR describes the fractional rate of stellar mass growth in a galaxy due to ongoing star formation. The 
sSFR has units of inverse time and galaxies with low sSFRs are said to have long star-formation time-scales.
Since the sSFRs of galaxies rises significantly from $z \sim 0$ to $z \sim 2-3$, and is best described by a power law $(1 + z)^{n}$
\citep[][]{Yoshida06,Karim11,Sobral13,Koyama13}.
We adopted a z-dependent sSFR
to separate actively SF galaxies from quenched galaxies, such that $\rm sSFR (z) = 10^{-10} \times (1+z)^{3}$ \citep{Koyama13}. SF galaxies are galaxies
with $\rm sSFR_{MAGPHYS} > sSFR (z)$, while $\rm sSFR < sSFR (z)$ 
would correspond to quiescent (or transitional) galaxies.

\begin{figure}
	\includegraphics[width=0.7\columnwidth,angle=90]{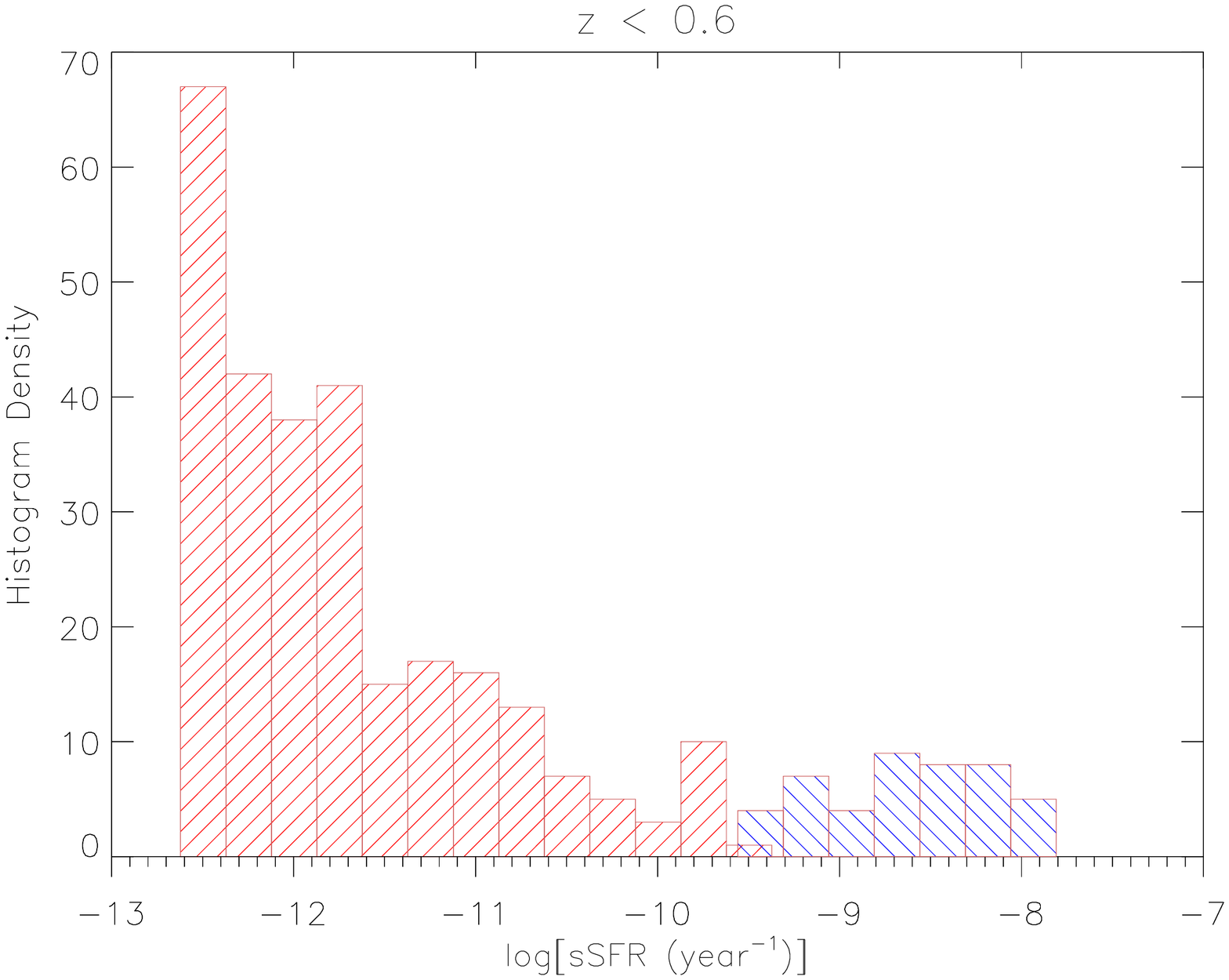}
	\includegraphics[width=0.7\columnwidth,angle=90]{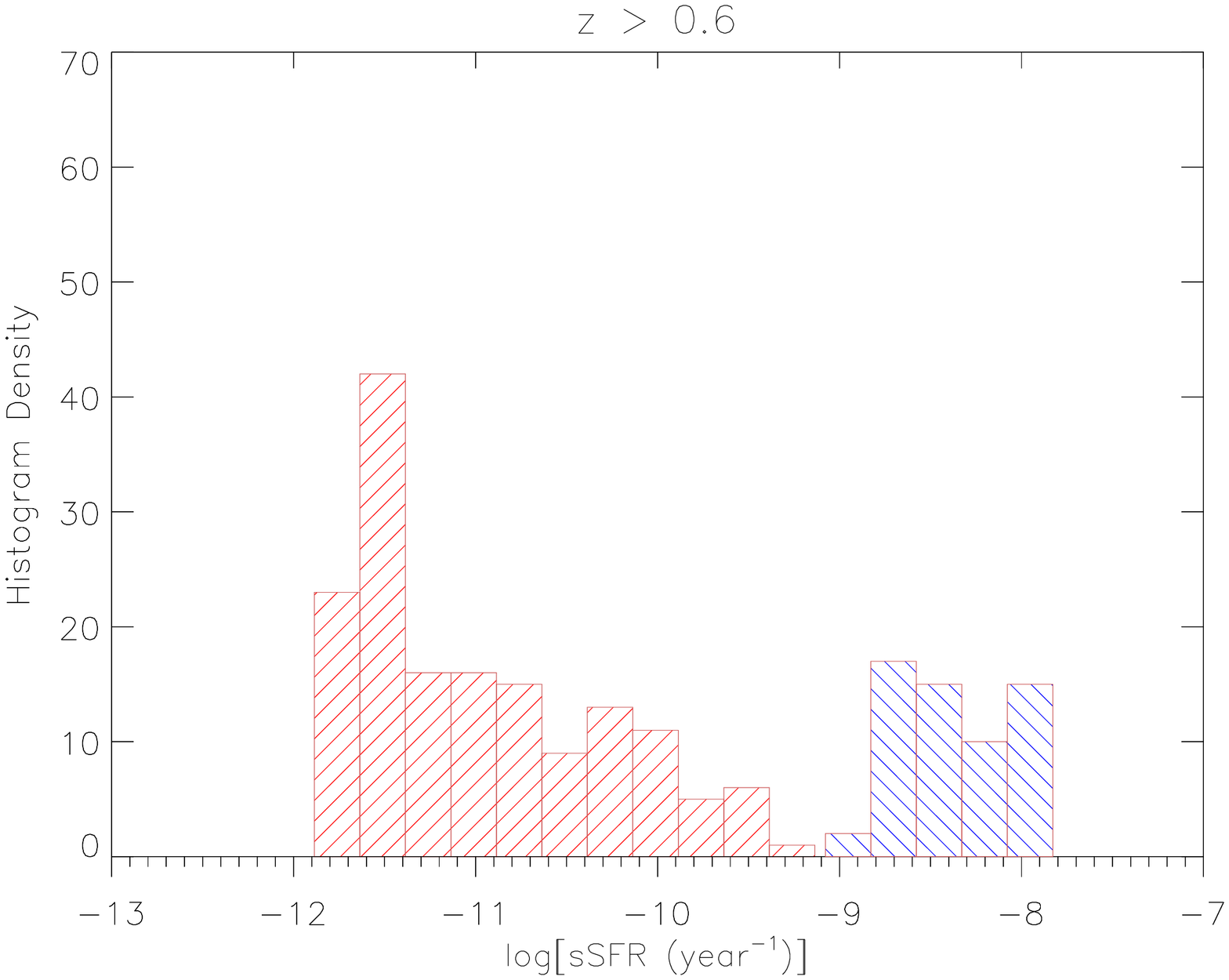}
    \caption{Histogram of the sSFR for the intermediate redshift bin (upper panel) and for the high redshift bin (lower panel). Red bar represent quiescent galaxies, while blue
    bars refer to SF galaxies.}
    \label{fig:histosSFR}
\end{figure}

The adoption of this sSFR(z) cut is reasonable if we look to the histogram distribution of sSFR for the two redshift bins (Fig.\ref{fig:histosSFR}).
 For $z < 0.6$ we see that there are already two peaks in the distribution (that correspond to quiescent and SF galaxies) and they becomes even more evident for
 the higher redshift bin ($z > 0.6$). Applying the K-S test, it shows that the probability that the sSFR for quiescent and SF galaxies are draw from different population 
is about 98\% for both redshift bins. The quiescent population represents 72\% of the sample at $z > 0.6$ increasing to 86\% for galaxies at $z < 0.6$,
giving additional support to the fact that  galaxies at higher redshifts have on average higher SFRs.

To analyse SFR/sSFR against projected radial distance (left panels) and against stellar mass (right panels), we adopted four bins of radial 
distance and stellar mass and we present the mean and standard deviation for each bin. 
In this way, we avoid that the trend is dominated by a specific region of the parameter space.
The blue triangles represent SF galaxies and  red circles quenched galaxies. This colour scheme will be used for all figures unless stated otherwise.

\begin{figure}
\center
	\includegraphics[width=0.55\columnwidth,angle=90]{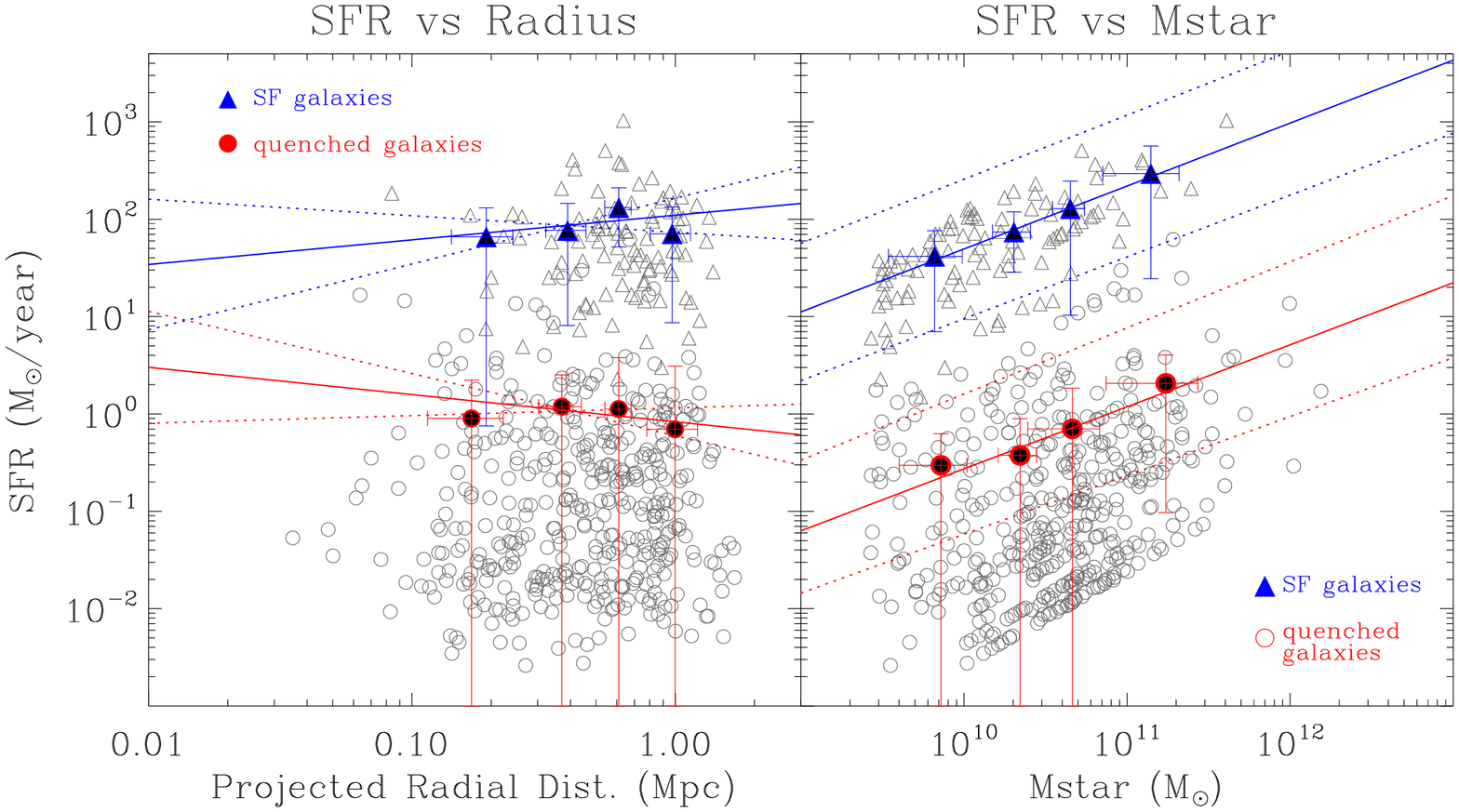}
    \caption{SFR against projected radial distance (left panel) and  $\rm SFR \times M_{\rm star}$ (right panel). 
    The grey triangles represent the SF population while the grey circles the quiescent galaxies.
    Blue triangles represent the mean for each adopted bin of actively star-forming galaxies and red circles represent 
    the mean for each adopted bin of quiescent/transient galaxies. 
    The blue solid line is the mean best fit for SFR x $M_{\rm star}$ for the star-forming galaxies,
    while the red solid line is the mean best fit for quiescent galaxies. Dashed lines are the one-sigma errors.}
    \label{fig:SFR_dist_Mstar}
\end{figure}

 In Fig.~\ref{fig:SFR_dist_Mstar} we show SFR $\times$ density (left panel) and SFR $\times M_{\rm star}$ (right panel). From this figure 
 we do not find any significant SFR-density correlation, suggesting that SF activity in the $0.4 < z < 0.9$ range does not significantly change 
within cluster environment.
Looking to the SFR-$M_{\rm star}$ relation, both populations show a strong correlation of the SFR on the stellar mass in the sense that 
more massive galaxies have higher star formation rates (with a Pearson correlation coefficient of 0.99 for the quiescent galaxies and 0.97 for  
the star-forming galaxies).
%suggesting that what drives SF is stellar mass. 

\begin{figure}
\center
	\includegraphics[width=0.55\columnwidth,angle=90]{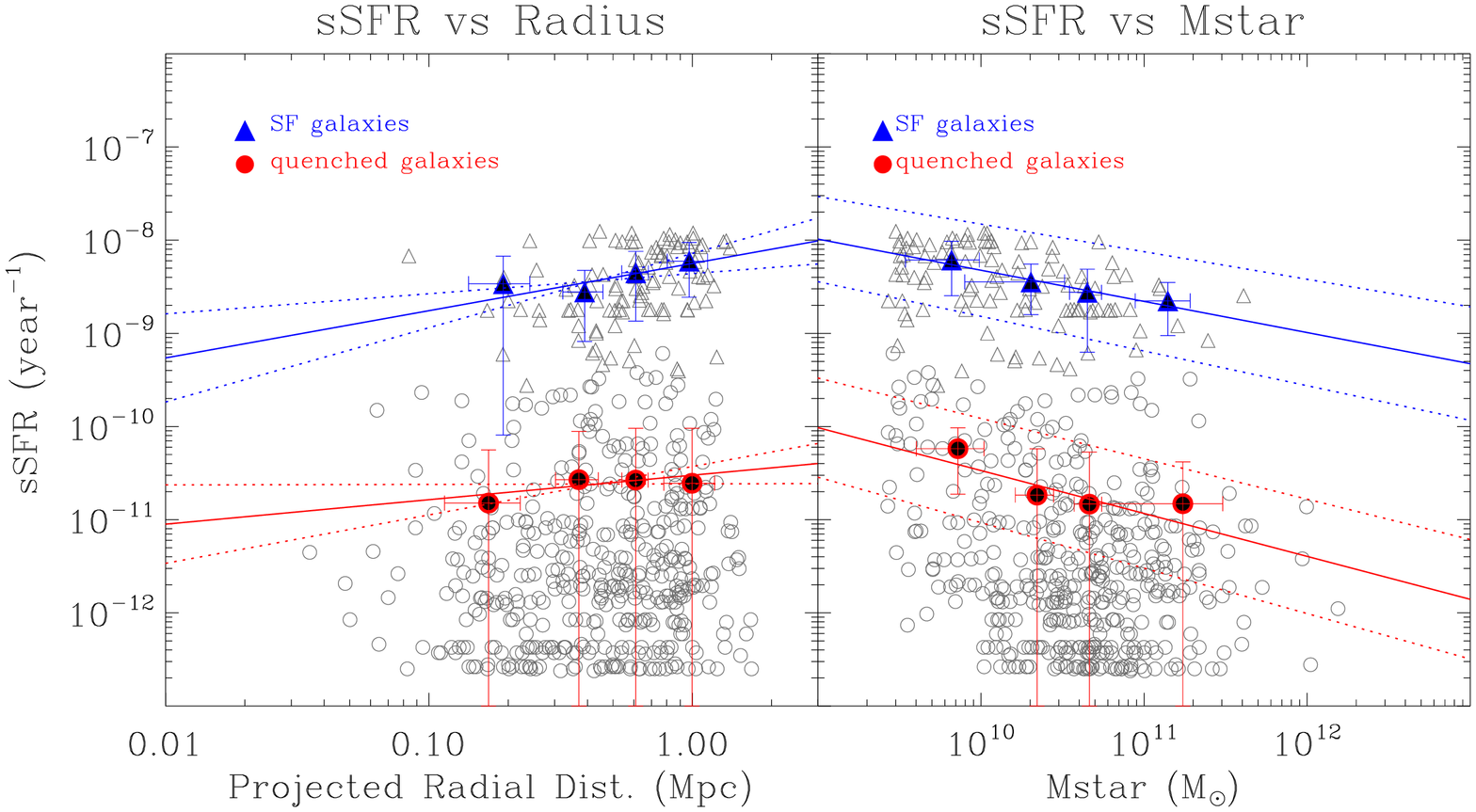}
    \caption{sSFR against projected radial distance (left panel) and  sSFR agains stellar mass (right panel). 
    The grey triangles represent the SF population while the grey circles the quiescent galaxies.
    Blue triangles represent  the mean for each adopted bin of  actively 
    star-forming galaxies and red circles represent  the mean for each adopted bin of quiescent/transient galaxies. On the left panel, 
    the solid lines are the mean best fits for $\rm sSFR \times$ projected radial distance and on the right panel the solid lines represent the 
   mean best fits for $\rm sSFR \times M_{\rm star}$. Solid blue-lines refer to the star-forming galaxies and red-lines to the quiescent galaxies.
   Dashed lines are the one-sigma errors.}
    \label{fig:ssfr}
\end{figure}

In Fig.~\ref{fig:ssfr} we present the sSFR against projected radial distance (left panel) and against stellar mass (right panel).
Our results show that the average sSFR may not follow an increasing trend with radius as suggested by 
\citet[e.g.,][]{Brodwin13}, indicating that SF activity does not significantly change within cluster environment.
We did not find any significant SFR-density or sSFR-density correlations 
(we could not reject the null hypothesis that there is 
no correlation between these two variables),
but we confirmed previous results in which low-mass galaxies have higher sSFR than higher-mass galaxies 
\citep{Feulner05,PerezGonzalez05,Zheng07,Noeske07b}. 
The Pearson correlation coefficient for sSFR-$M_{\rm star}$ relation is  -0.97 and -0.83  for the SF and quiescent galaxy population, respectively.

%meaning that more massive galaxies are forming less stars per unit of mass 

Since the correlations between SFR/sSFR and $M_{\rm star}$ are strong for both SF and quiescent galaxies, this may point to the fact that their 
properties are primarily determined by their stellar mass, and not by the environment they reside \citep[as suggested by][]{Muzzin12,Sobral11}.

\section{Discussion}
\label{disc}

To better analyze the SFR and sSFR dependence on both $M_{\rm star}$ and on environment through time, we divided our sample into two bins of redshift: an intermediate-redshift bin 
(cluster galaxies at $z < 0.6$) and a high-redshift bin (cluster galaxies at $z > 0.6$). The redshift of $z=0.6$ was chosen as the redshift that divided our sample
into two sub-samples with approximately the same number of galaxies. 

\begin{figure}
	\includegraphics[width=0.55\columnwidth,angle=90]{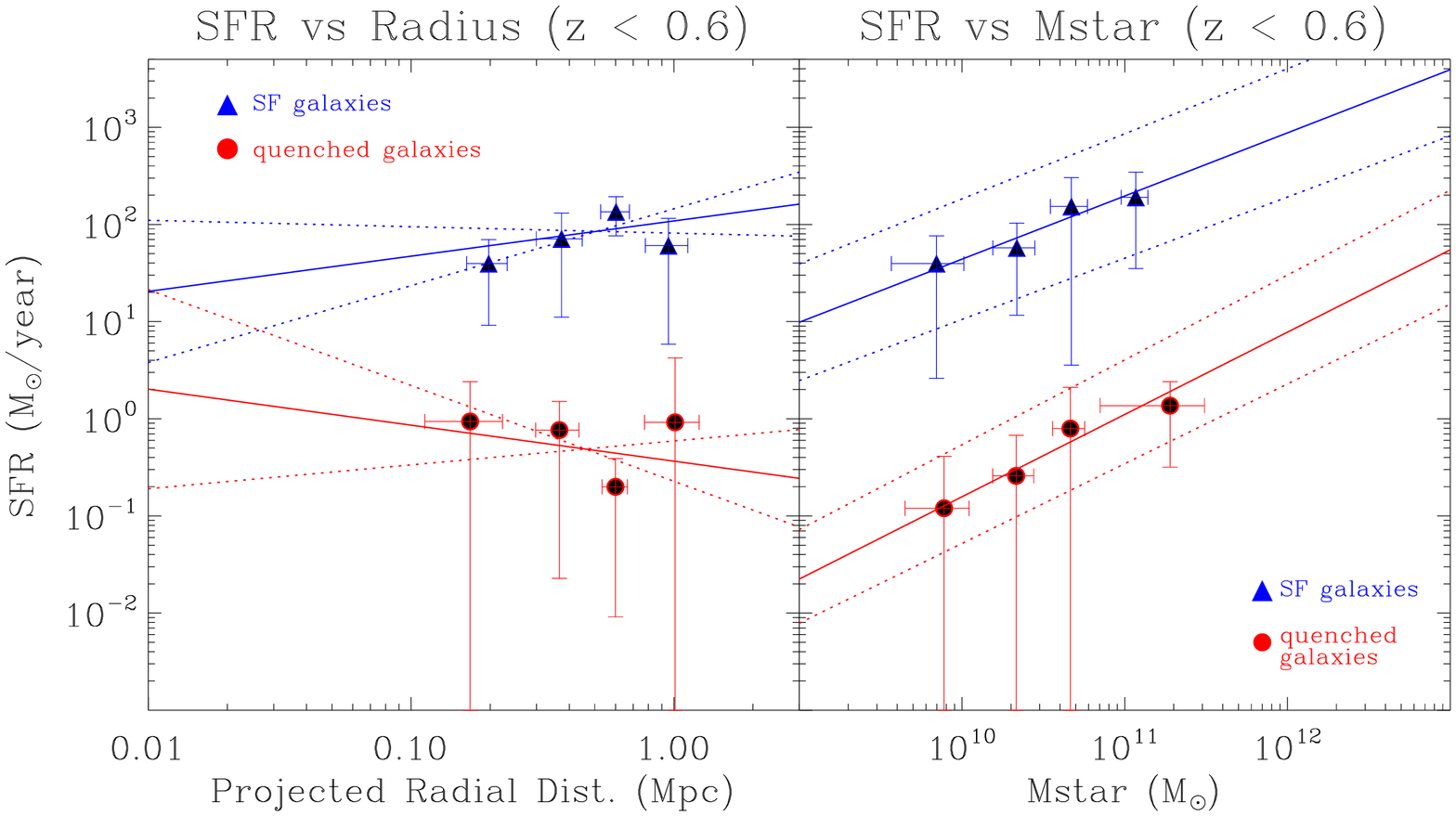} 
	\par
	\vspace{0.5cm}
	\includegraphics[width=0.55\columnwidth,angle=90]{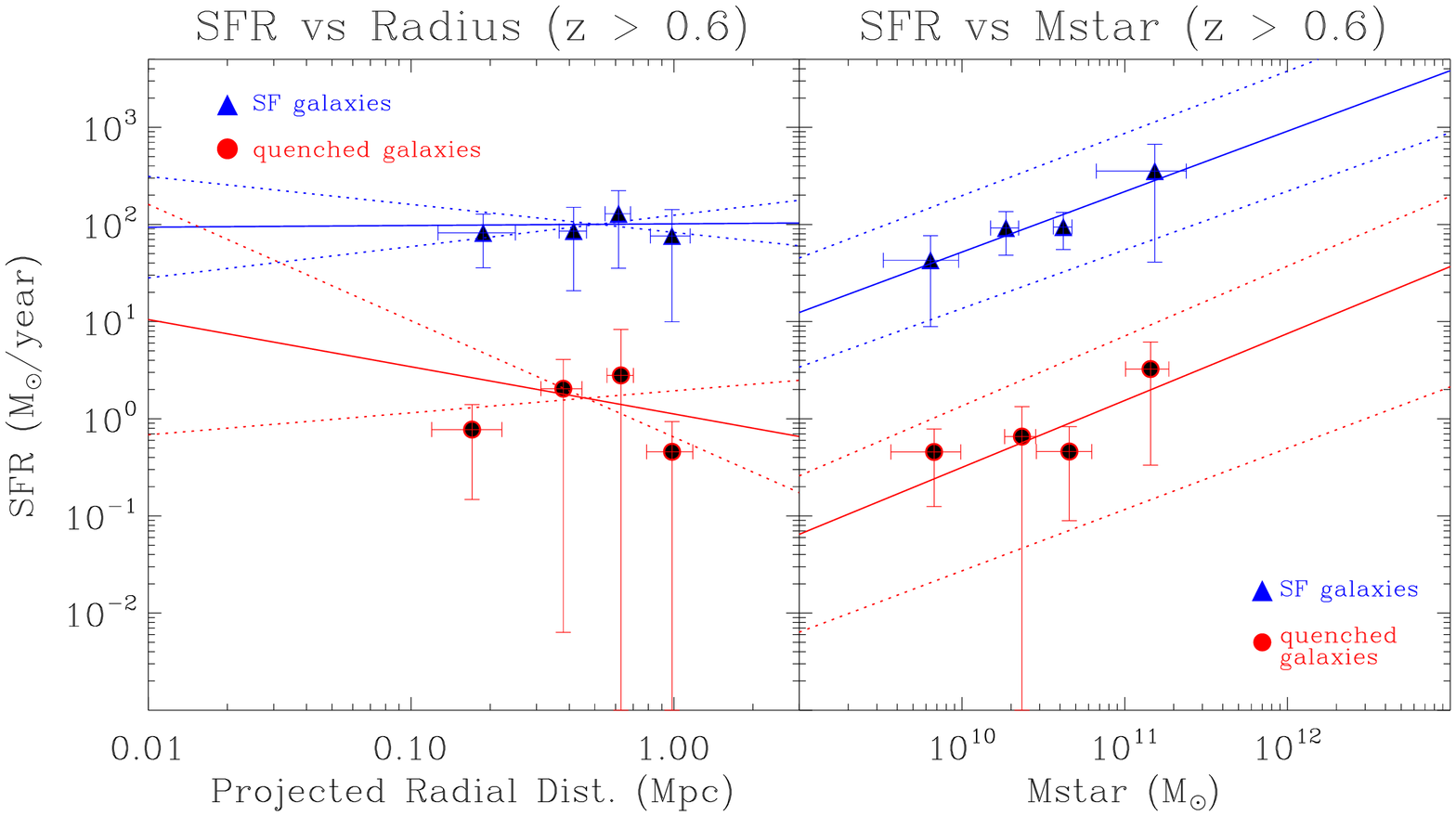}
    \caption{SFR against projected radial distance (left panels) and  SFR agains stellar mass (right panels), for two redshift bins: upper panels we show galaxies
    at redshift $z < 0.6$, and lower panels galaxies with $z > 0.6$. Blue triangles represent the mean for each adopted bin of actively star-forming galaxies 
    and red circles represent the mean for each adopted bin of quiescent galaxies. On the left panels, 
    the solid lines are the mean best fit, where the solid blue-lines refer to the star-forming galaxies and red-lines to the quiescent galaxies. Dashed lines are the one-sigma errors.}
    \label{fig:SFRz}
\end{figure}

In Fig.~\ref{fig:SFRz} we show the SFR against projected radius (left panels) and against stellar mass (right panels) for cluster galaxies at $z < 0.6$ (upper panels) and
cluster galaxies at $z > 0.6$ (lower panels). 
Taking into account the 1-$\sigma$ errors, there is no clear correlation between SFR and projected radius. 
Thus, our results show that within the cluster environment, SFR and density does not correlate (in any of the redshift bins analysed),
and it may be that the effects of environment on the galaxy 
population at higher redshifts are relatively minor in all environments except for the richest galaxy clusters, 
something that has been suggested by \citet{Sobral11} and \citet{Muzzin12}.

Also from Fig.~\ref{fig:SFRz}, our results show a strong dependence of the SFR on the stellar mass similar to what is found for field galaxies up to
redshift $\sim$ 1 \citep{LaraLopez10,Peng10}. The mass dependence of the SFR for SF galaxies is not surprising as many studies have stablished a tight 
correlation between SFR and $M_{\rm star}$, that is, the main sequence of SF galaxies \citep{Brinchmann04,Daddi07,Karim11,Sobral14}.
However, \citet{Noeske07} found that the dependence of the SFR on $M_{\rm star}$ can be seen for the galaxies with
reliable signs of star formation (main sequence of SF galaxies), while galaxies with no evident signs of SF  form  an horizontal sequence on 
the $\rm SFR-M_{\rm star}$ diagram. In stark contrast, our results reveal that the quiescent population also shows a dependence of the SFR on $M_{\rm star}$.

\begin{figure}
	\includegraphics[width=0.55\columnwidth,angle=90]{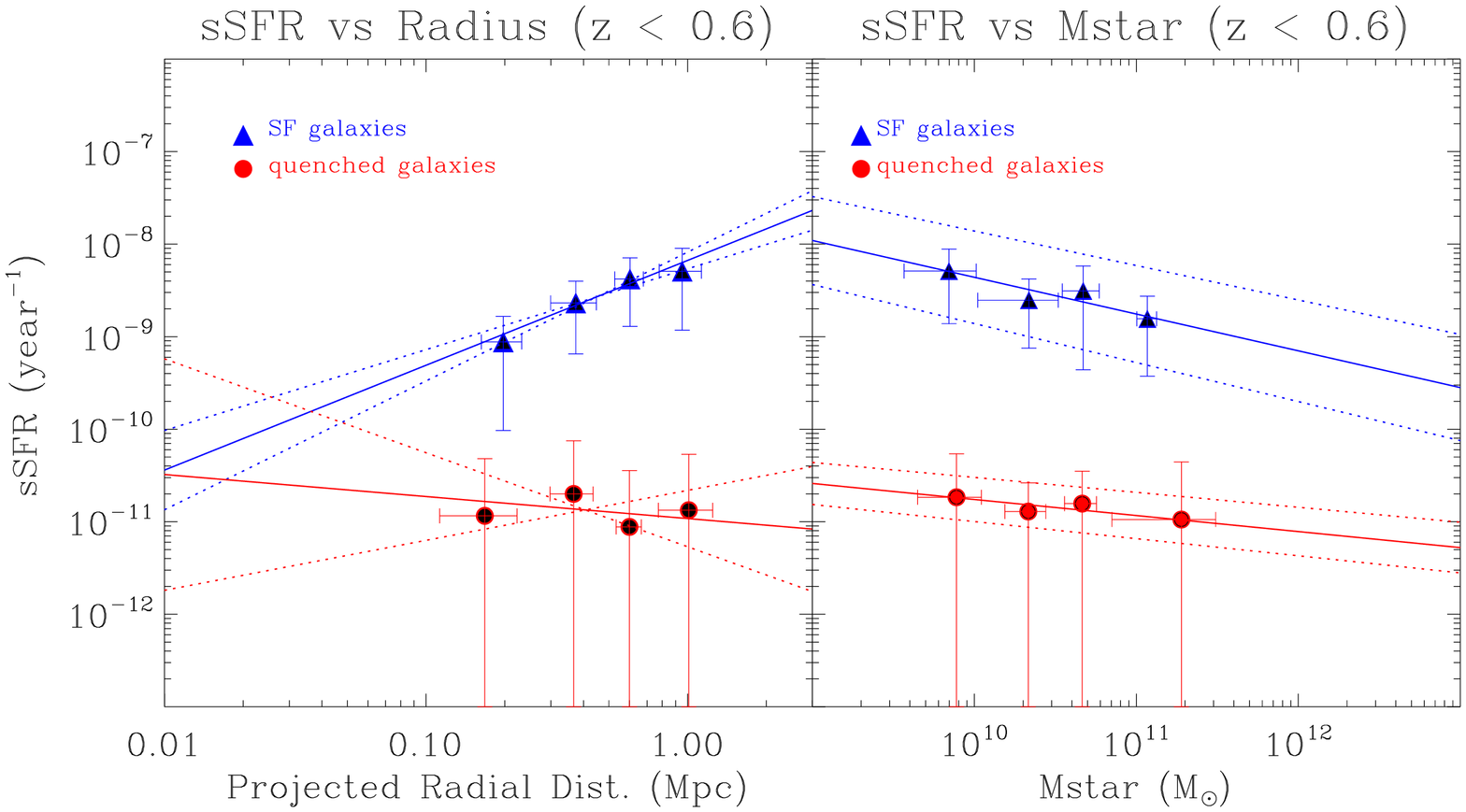}
	\par
	\vspace{0.5cm}
	\includegraphics[width=0.55\columnwidth,angle=90]{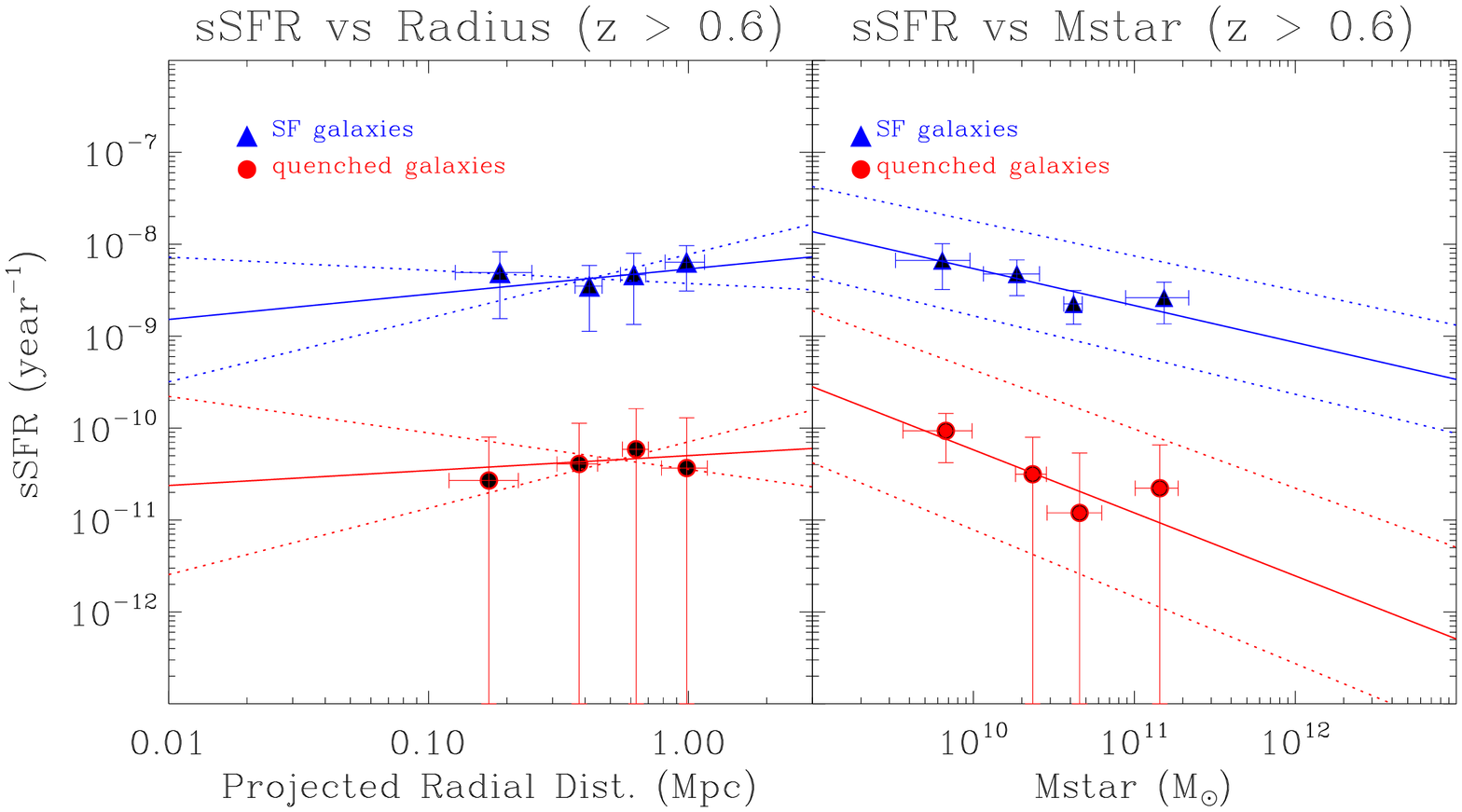}
    \caption{sSFR against projected radial distance (left panels) and  sSFR agains stellar mass (right panels), for two redshift bins: upper panels we show galaxies
    at redshift $z < 0.6$, and lower panels galaxies with $z > 0.6$. 
    Blue triangles represent the mean for each adopted bin of actively 
    star-forming galaxies and red circles represent the mean for each adopted bin of quiescent/transient galaxies.  
    On the left panels, 
    the solid lines are the mean best fit, where the solid blue-lines refer to the star-forming galaxies and red-lines to the quiescent galaxies. Dashed lines are the one-sigma errors.}
    \label{fig:sSFRz}
\end{figure}

In Fig.~\ref{fig:sSFRz} we show the sSFR against projected radius (left panels) and against stellar mass (right panels) for cluster galaxies at $z < 0.6$ (upper panels) 
and cluster galaxies at $z > 0.6$ (lower panels). 

Except for the SF galaxies at $z < 0.6$, there is no correlation between sSFR and projected radius. 
Recent works have obtained that, at least for SF galaxies, there is no significant SFR-density or sSFR-density correlation \citep{McGee11,Peng10,Tadaki12,Muzzin12}.
The results drawn from our current work are consistent with those previous findings not only for the SF population but also for quenched galaxies. 
We caution that the flat SFR/sSFR-density is valid within cluster environment and may not hold for higher redshifts. Also, it is worth mentioning that our results 
represent mean values for SFR/sSFR averaged over galaxies of different clusters possibly with different dynamical states.

The sSFR-density correlation found for the SF galaxies at $z < 0.6$, might be due to merging effects.
 This effect is because the merging is predicted to be a gradual rather than one time catastrophic event, and hydrodynamical simulations 
 show that this gradual build up began as 
early $z \sim 2$ \citep[e.g.,][]{Mundy17, Martin17, Feldmann17} but why or how the effects are most noticeable below $z \sim 0.6$ 
requires further work that is beyond the scope of this paper.

Also form Fig.~\ref{fig:sSFRz} we see an anti-correlation between the sSFR and $M_{\rm star}$, independent of the redshift bin
analysed, showing that lows-mass galaxies have higher sSFR than higher-mass galaxies.
These results show not only that more massive galaxies are forming less stars per unit of mass,
but it supports a scenario in which massive galaxies formed most of their stars earlier and on shorter timescales, while less-massive galaxies evolve on longer 
timescales \citep[``downsizing'',][]{Popesso11,Sobral11,Scodeggio09}.

\section{Summary and Conclusions}
\label{conc}

We analysed 17  galaxy clusters, and investigated, for the first time, the dependence of the SFR
and sSFR as a function of projected distance (as ameasure of the galaxy density environment) and stellar mass for cluster galaxies in
 an intermediate-to-high redshift range ($ 0.4 < z < 0.9$). 
We used up to nine flux points (BVRIZYJH$K_{s}$ magnitudes), its errors and redshifts to compute $M_{\rm star}$, SFR and sSFR through 
spectral energy distribution
fitting technique. 
To separate our galaxies in SF and quiescent population we adopted a specific star-formation rate as a function of redshift, 
$sSFR = 10^{-10} \times (1+z)$, and classified as actively SF galaxies
the ones with sSFR above this threshold, while the ones with lower values belong to the quiescent population.

To analyse the SFR and sSFR history we split our sample in two redshift bins: galaxies at $z < 0.6$ and $z > 0.6$.  
We separate the effects of environment and stellar mass on galaxies by comparing the properties of star-forming and quiescent galaxies at fixed environment 
(projected radius) and fixed stellar mass.

An observational challenge here was to test if the ``universality'' of the main sequence holds at an intermediate-to-high redshift range, where 
global star-formation activity is higher. We confirmed the existence of a universal galaxy main sequence in clusters. Plus, we also showed
that for both, SF and quiescent population, SFR correlates with
stellar mass. We also found an anti-correlation between the sSFR and $M_{\rm star}$, independent of the redshift bin
analysed, showing that lows-mass galaxies have higher sSFR than higher-mass galaxies. These results show not only that more massive galaxies 
in clusters
are forming less stars per unit of mass,
but our results support  a scenario in which massive galaxies formed most of their stars earlier and on shorter timescales, while less-massive galaxies evolve on longer 
timescales (``downsizing'').

From our results, we did not find any significant SFR-density or sSFR-density correlations (we could not reject the null hypothesis that there is 
no correlation between these two variables), suggesting that
SF activity does not significantly change within cluster environment, making it evident that mass is the parameter that drives SFR \citep[in line with previous finding of][]{Muzzin12,Koyama13,Darvish15,Darvish16}.

%We could not draw firm conclusions on the SFR-density relation due to the low correlations found. It may be that the effects of environment on the galaxy 
%population at higher redshifts are relatively minor in all environments except for the richest galaxy clusters, something that has been suggested by \citet{Sobral11} and \citet{Muzzin12}.

\section*{Acknowledgements}
We would like to thank the anonymous referee for valuable and constructive comments that improved the quality of this work.
The authors also thank F. Durret and N. Martinet for making available the cluster galaxies data in an appropriate table,
E. da Cunha for clarifying  MAGPHYS applications and, L. Martins and P. Coelho for fruitful discussions.
T. F. L acknowledges FAPES and CNPq for financial support (grants: 2012/00578-0 and 303278/2015-3, respectively),
and thanks Northwestern University Center for Interdisciplinary Exploration and Research in Astrophysics (CIERA) for hosting while we carried out part of this research.

%%%%%%%%%%%%%%%%%%%%%%%%%%%%%%%%%%%%%%%%%%%%%%%%%%



%\bibliographystyle{mnras}


\bibliographystyle{mnras} 
\include{adsjournalnames} 
\bibliography{refs}

%%%%%%%%%%%%%%%%%%%%%%%%%%%%%%%%%%%%%%%%%%%%%%%%%%

%%%%%%%%%%%%%%%%% APPENDICES %%%%%%%%%%%%%%%%%%%%%

\appendix

\section{Completeness}
\label{Mstarz}

To compare the stellar masses in this sample, we defined the completeness stellar mass as a function of redshift 
following Section 3.1 of \citet{Lagana13} \citep[but also adopted in][]{Bolzonella10,Pozzetti10}. 
This is the lowest mass at which the galaxy stellar mass function can be considered as reliable and unaffected by incompleteness. 

For each galaxy, we computed the ``limiting mass'', which is the stellar mass that this galaxy would have if its apparent magnitude was
equal to the sample limit magnitude (i.e., I = 23.2): $\log(\mathcal{M}^{\rm star}_{\rm lim}) = \log(\mathcal{M}) + 0.4 \times (I-23.2)$, where $\mathcal{M}$ is the stellar mass of the 
galaxy derived from MAGPHYS with apparent magnitude I.

We devided our sample in for redshift bins and, for each bin, we computed the 20\% faintest galaxies (grey points in Fig. A1) and then,
for each redshift bin we define the value corresponding to 95\% of the distribution of limiting masses as a minimum mass. 
Fitting these four bins of limiting mass values, we have a mass completeness function independent of redshift, given by $\log(M^{\rm star}_{\rm lim}= 9.4$. We thus adopted this value as the 
lowest galaxy stellar mass that will be considered in our analysis. Also, we applied a K-S test and it shows that the probability that the mass 
distribution from the two redshift bins ($z < 0.6$ and $z > 0.6$) 
are draw from different population is about 15\% and do not introduce any bias in the results.

\begin{figure}
	\includegraphics[width=0.8\columnwidth,angle=90]{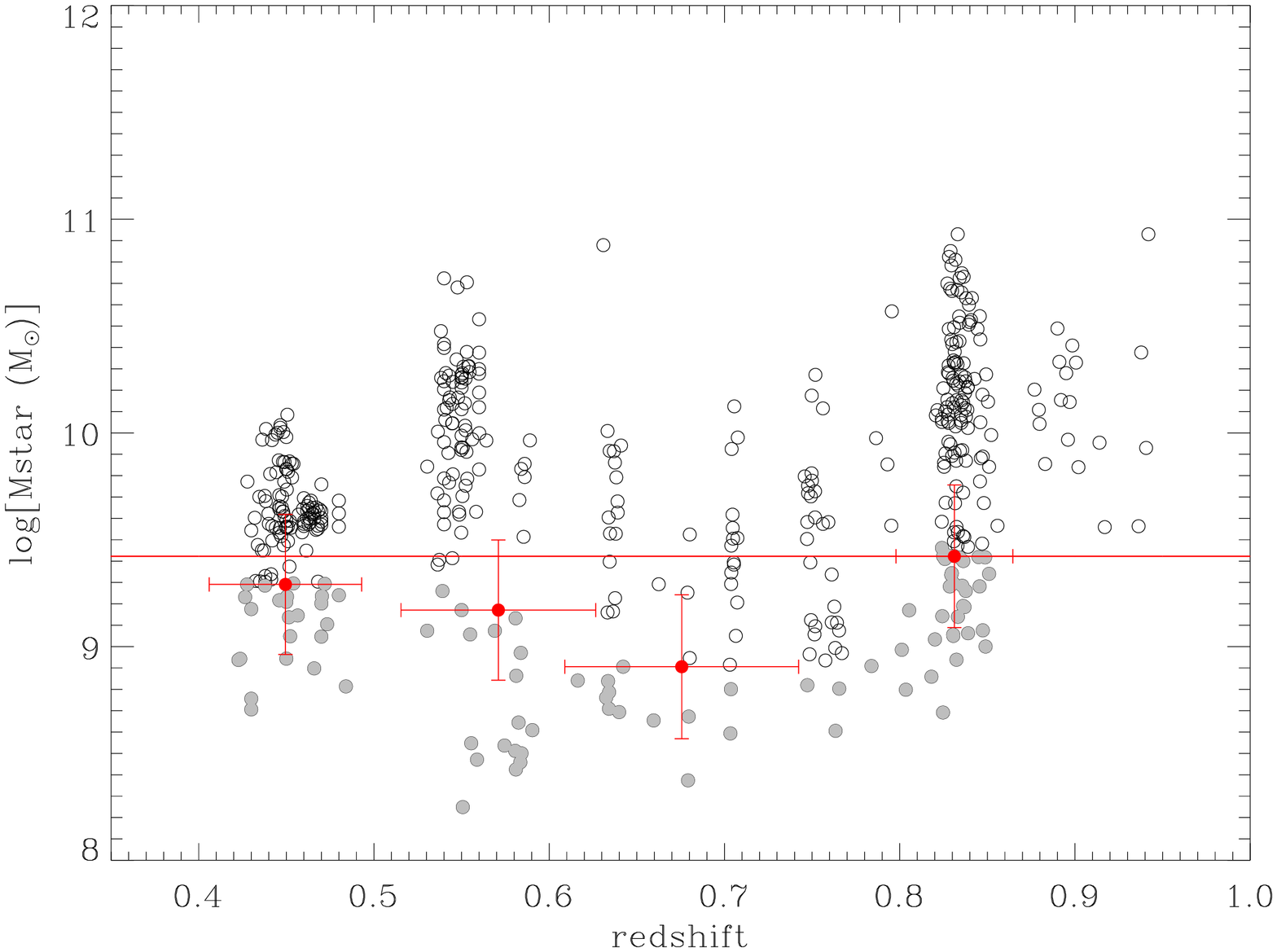}
    \caption{Stellar mass as a function of redshift for all galaxies analysed in this work.}
    \label{fig:Massalimite}    
\end{figure}

\section{SED fits and NIR data}
\label{ApendixSED}

For five clusters in our sample (LCDCS130, LCDCS173, LCDCS504, LCDCS531 and, LCDCS853) there are IRAC1 and IRAC2 available data.
Thus, in order to test how these bands would affect our results, we show in Fig. B1 the stellar mass derived with and without Spitzer bands.
Since we use more than one near infra-red data to constrain the IR regime, we see that the results are consistent and not including IRAC bands
does not affect the derived results because MAGPHYS uses a Bayesian approach to determine the $M_{\rm star}$ and SFR that takes into account uncertainties 
due to the lack of data in certain spectral ranges and/or degeneracies between physical parameters. 

\begin{figure}
	\includegraphics[width=0.8\columnwidth,angle=90]{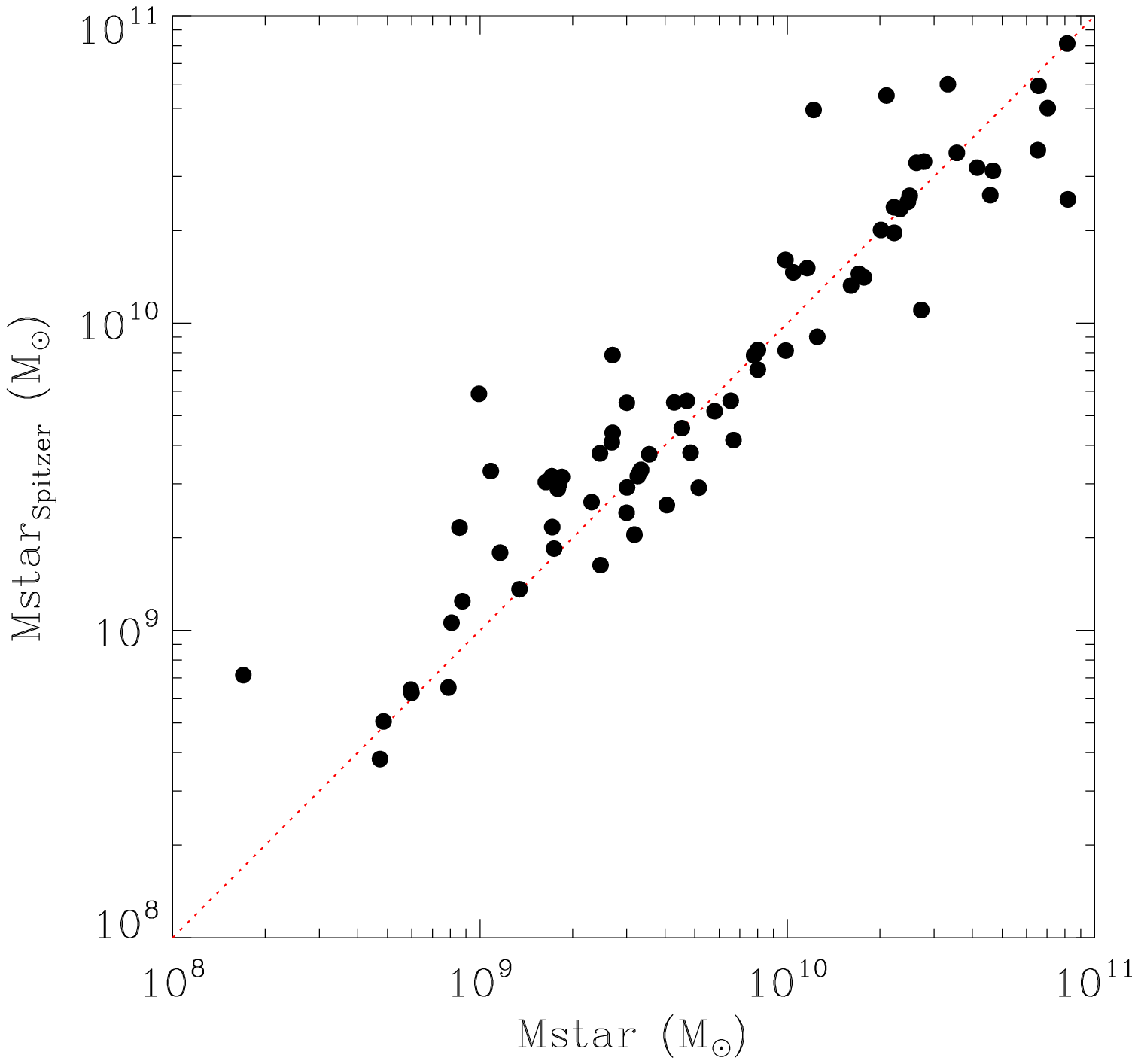}
	\includegraphics[width=0.8\columnwidth,angle=90]{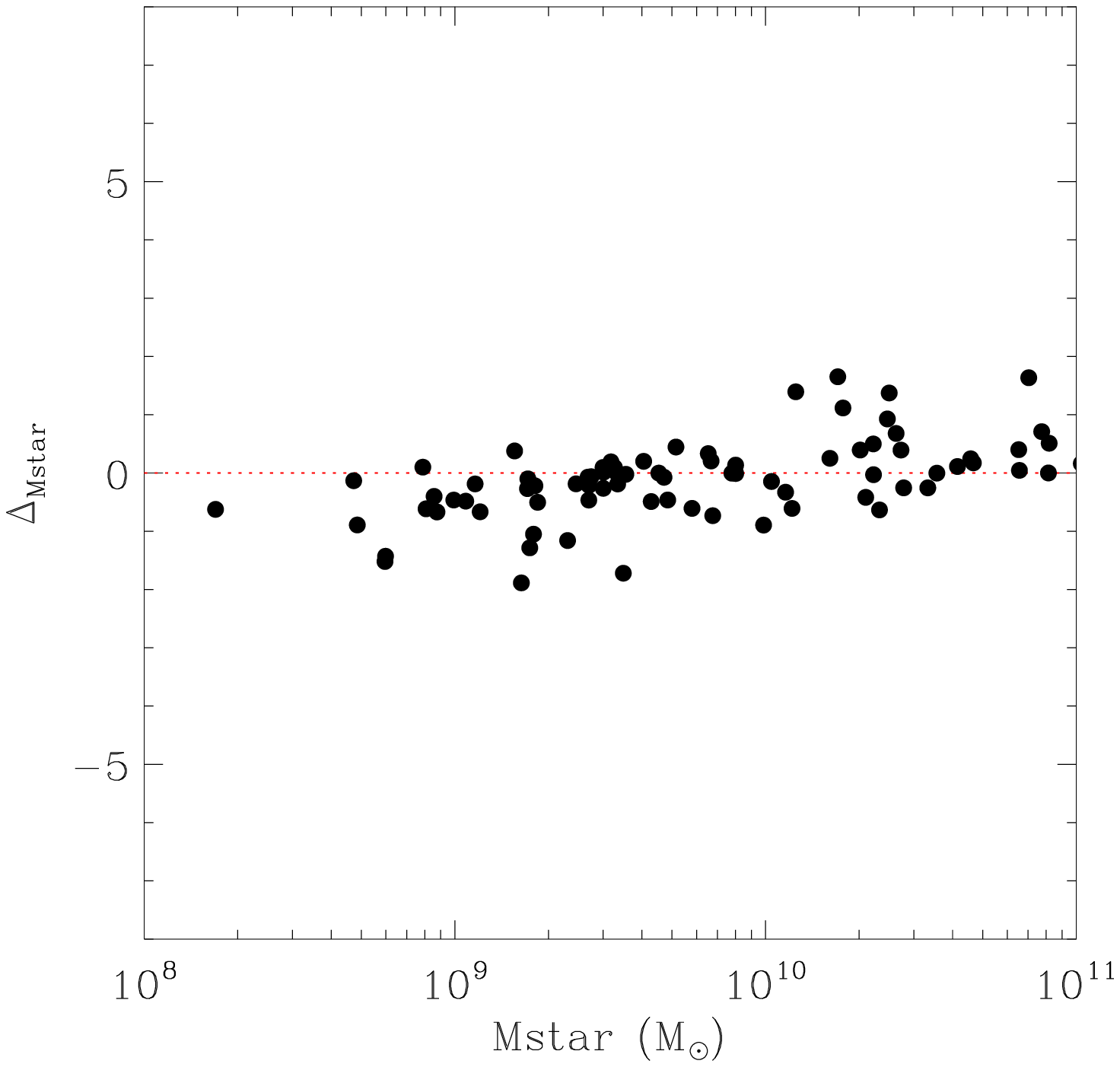}
    \caption{\textit{Upper panel}: Stellar mass computed  with Spitzer data as a function of $M_{\rm star}$ without Spitzer data. \textit{Lower panel}: logarithmic difference in mass ($\log(M_{\rm star}) - \log(M_{\rm star_{\rm Spitzer}}$)) as 
    a function of  $M_{\rm star}$ and line of zero offset between them in red.}
    \label{fig:mstarspitzer}
\end{figure}

Here, we also show some SED fits for typical ``quenched'' and SF galaxies at low (Fig. B2) and high redshifts (Fig. B3).
The best fit model is represented by the black solid lines that are fitted to the observed SED (red points). The blue lines correspond to the unattenuated stellar population spectrum.
The minor panels show the likelihood distribution of the output parameters derived from fits to the observed spectral energy distribution.

\begin{figure*}
	\includegraphics[width=0.95\columnwidth]{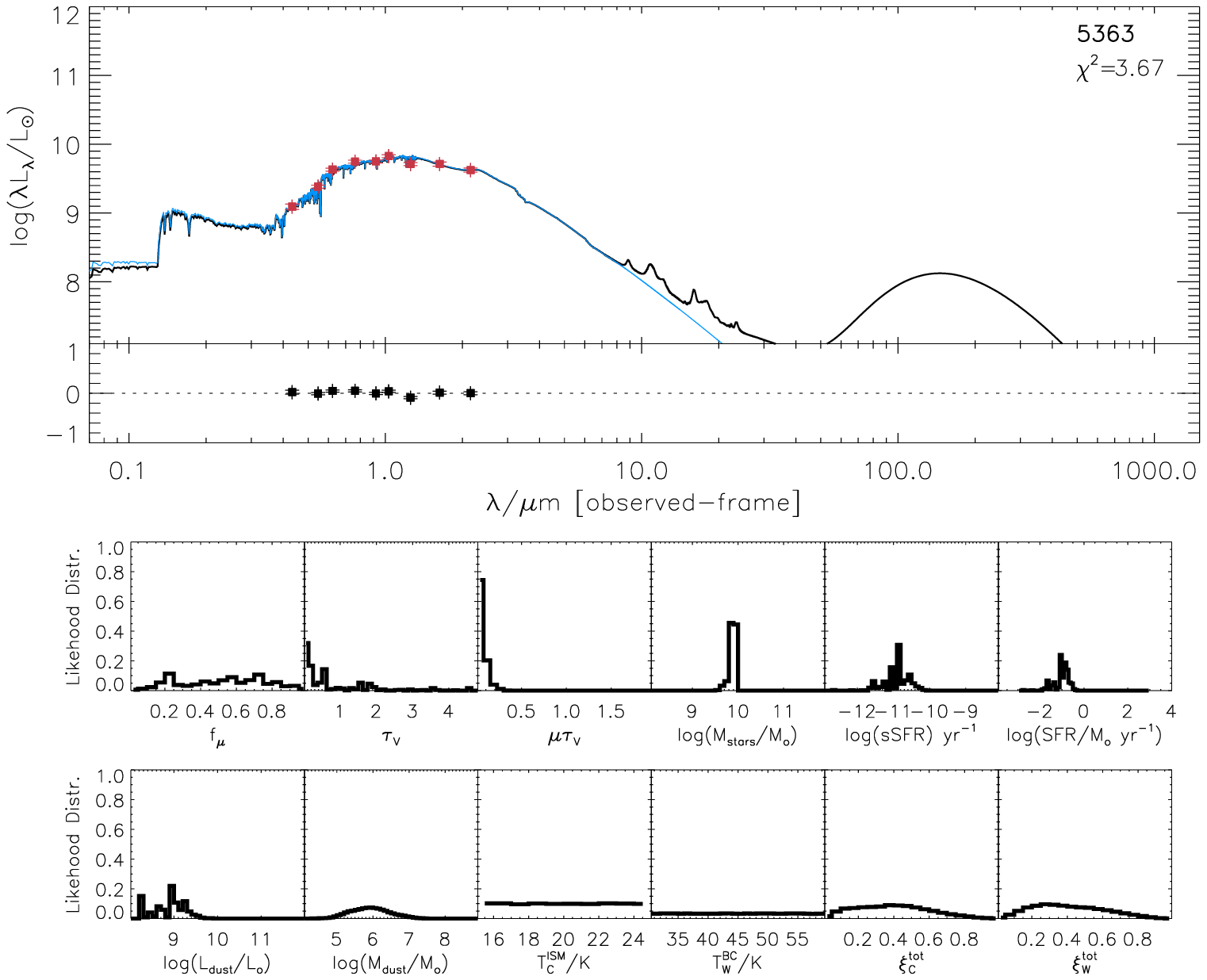}
	\includegraphics[width=0.95\columnwidth]{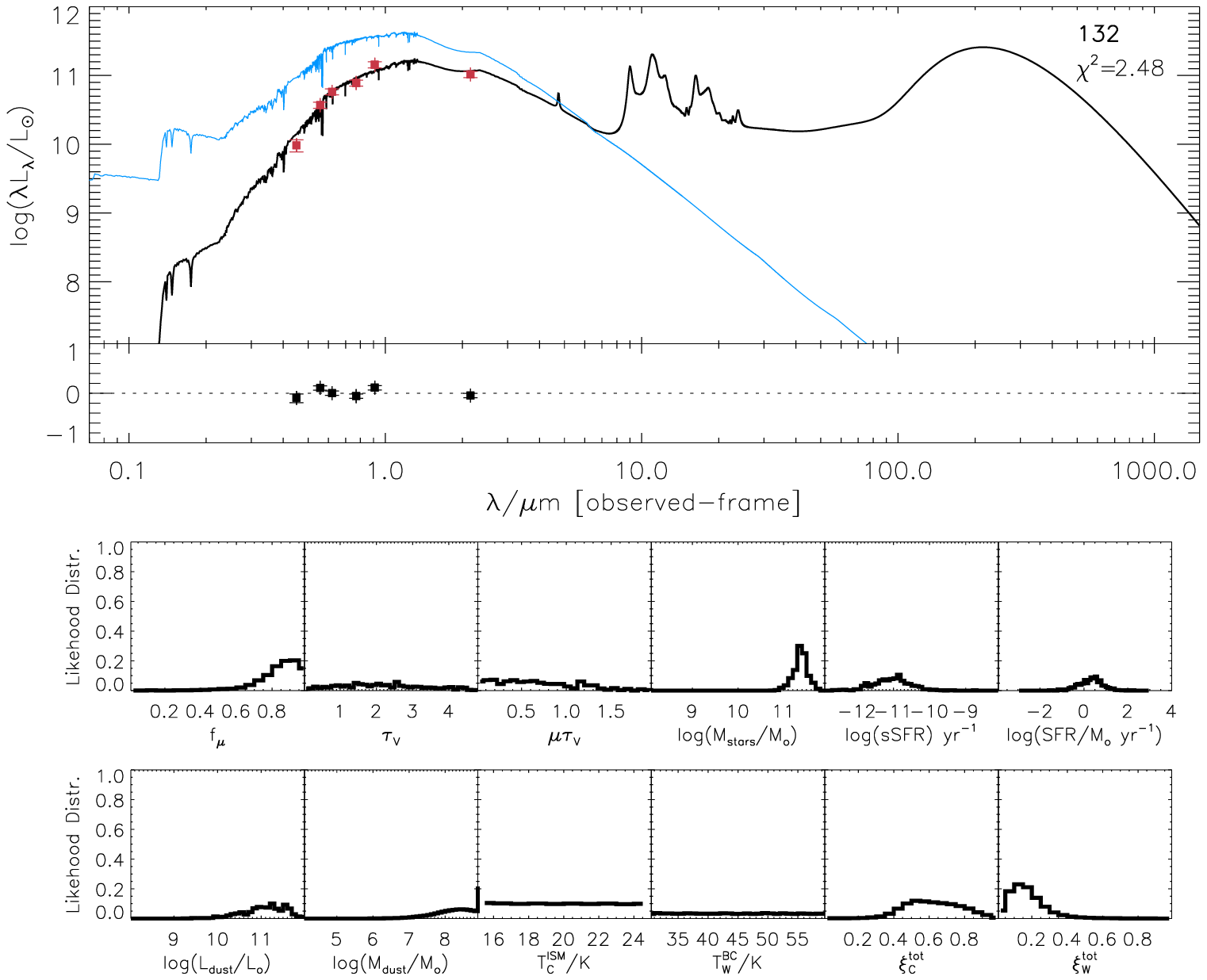}
	\vspace{0.5cm}
	\includegraphics[width=0.95\columnwidth]{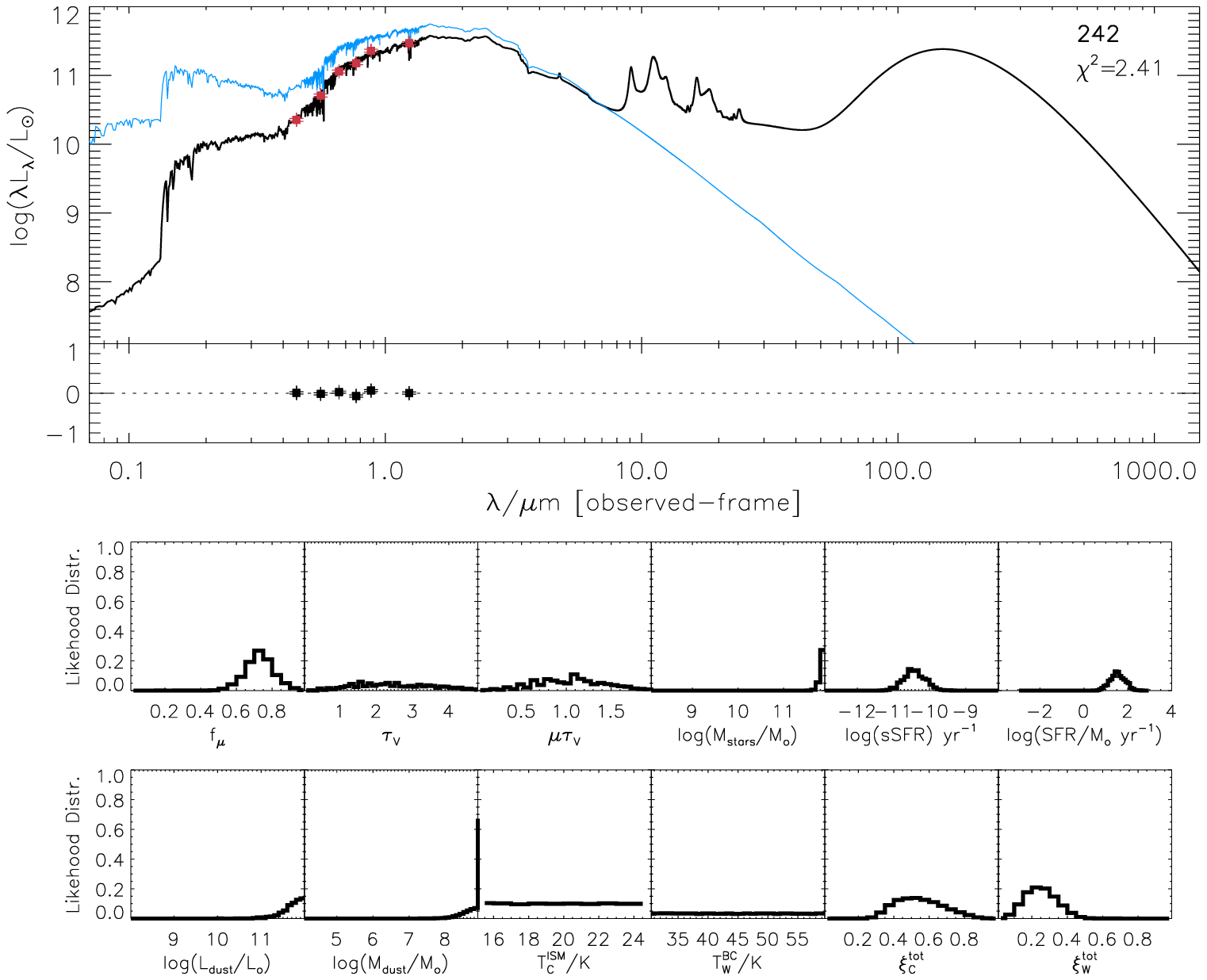}
	\includegraphics[width=0.95\columnwidth]{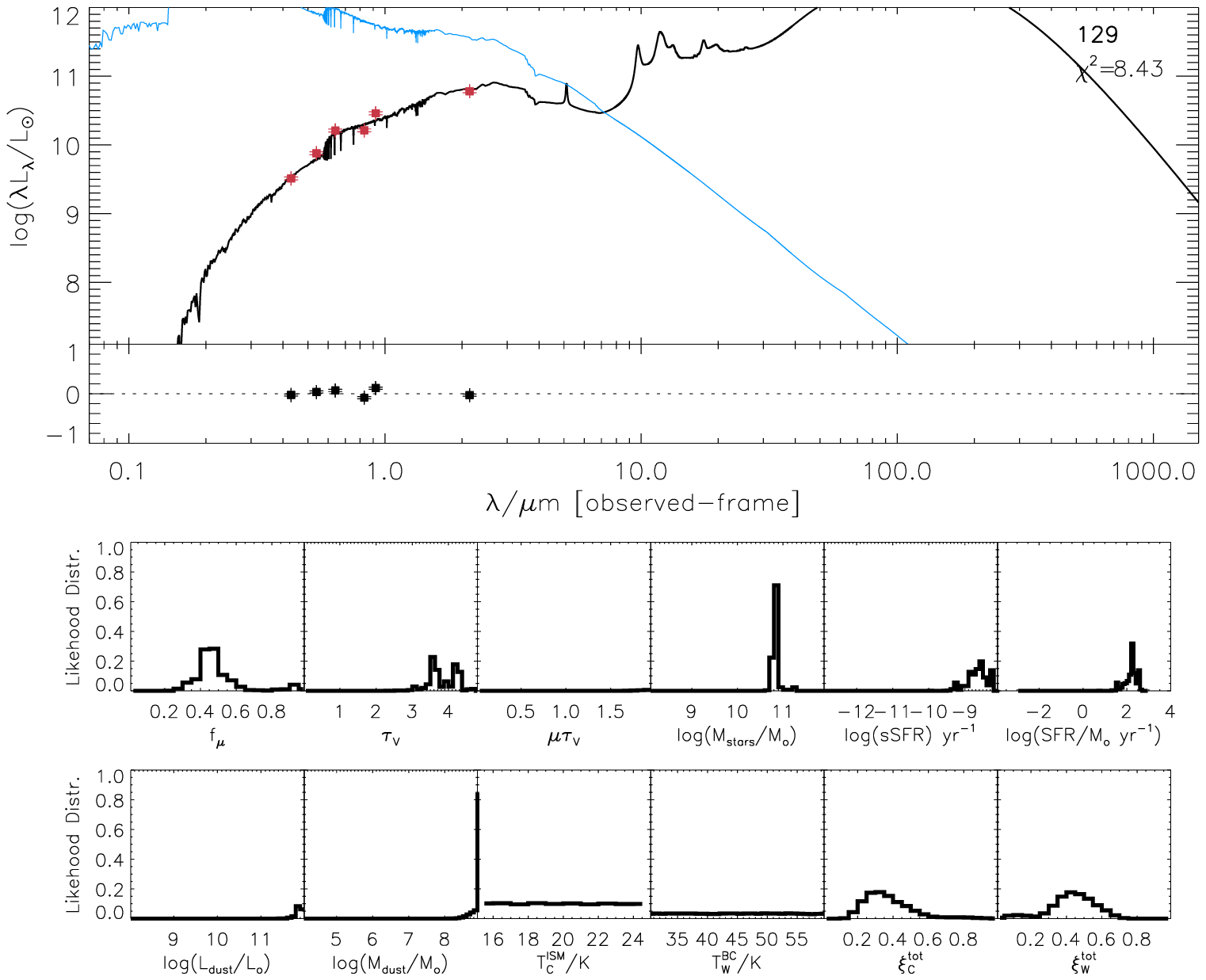}
    \caption{SED fits for low redshift galaxies. {Upper panels}:  typical  ``quenched'' galaxies from Abell 851 and RXC J1206.2-0848 
    {Lower panels}: typical SF galaxies from LCDCS 0829 and CL 0016+1609. Best model fits (black lines) to the observed spectral energy distribution (red points) of the galaxies. In each panel,
    the blue solid line is the the unattenuated stellar population spectrum. The minor panels show the likelihood distribution of the output parameters derived from fits to the observed spectral energy distribution.}
    \label{fig:SEDlowz}
\end{figure*}
    
\begin{figure*}
	\includegraphics[width=0.95\columnwidth]{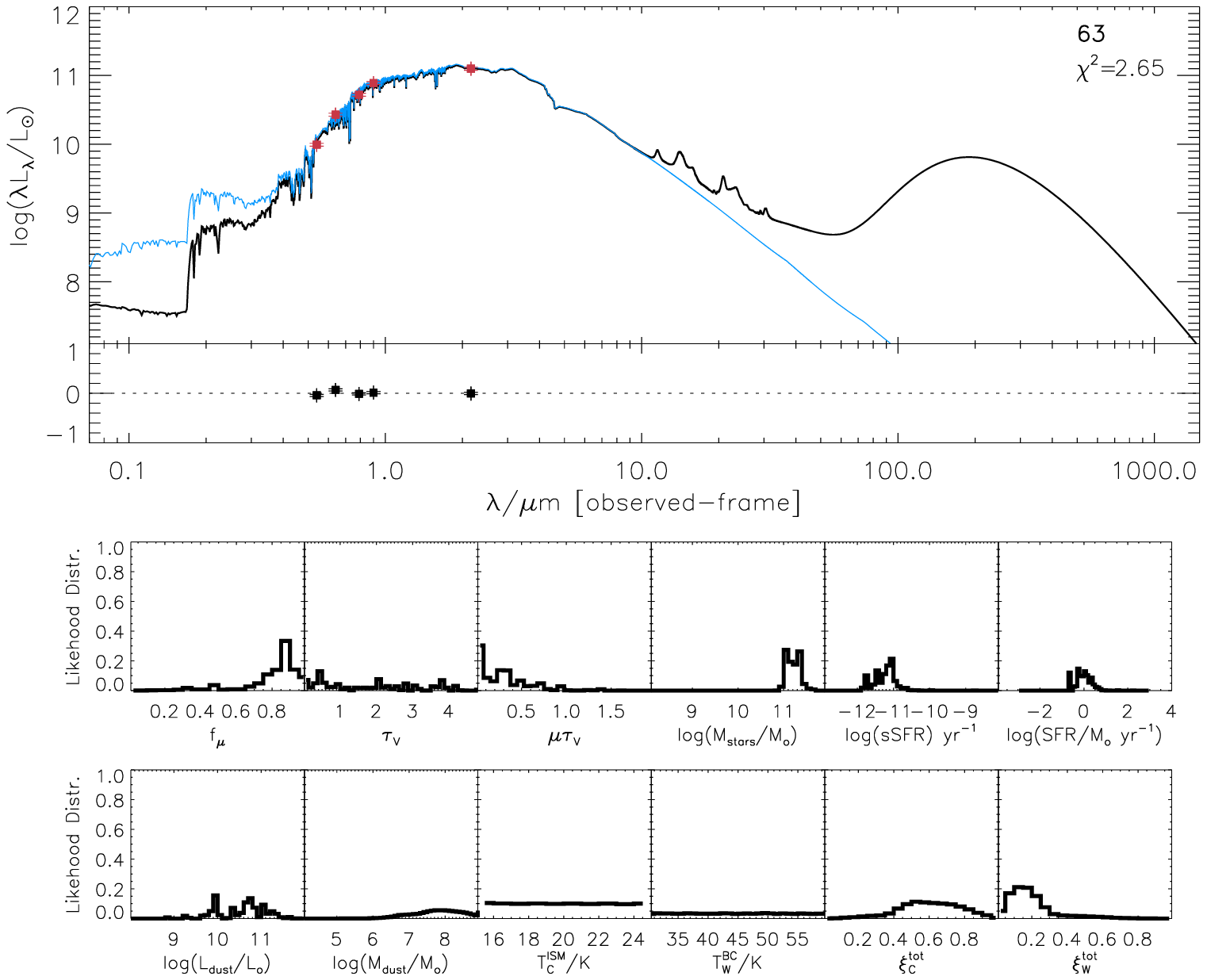}
	\includegraphics[width=0.95\columnwidth]{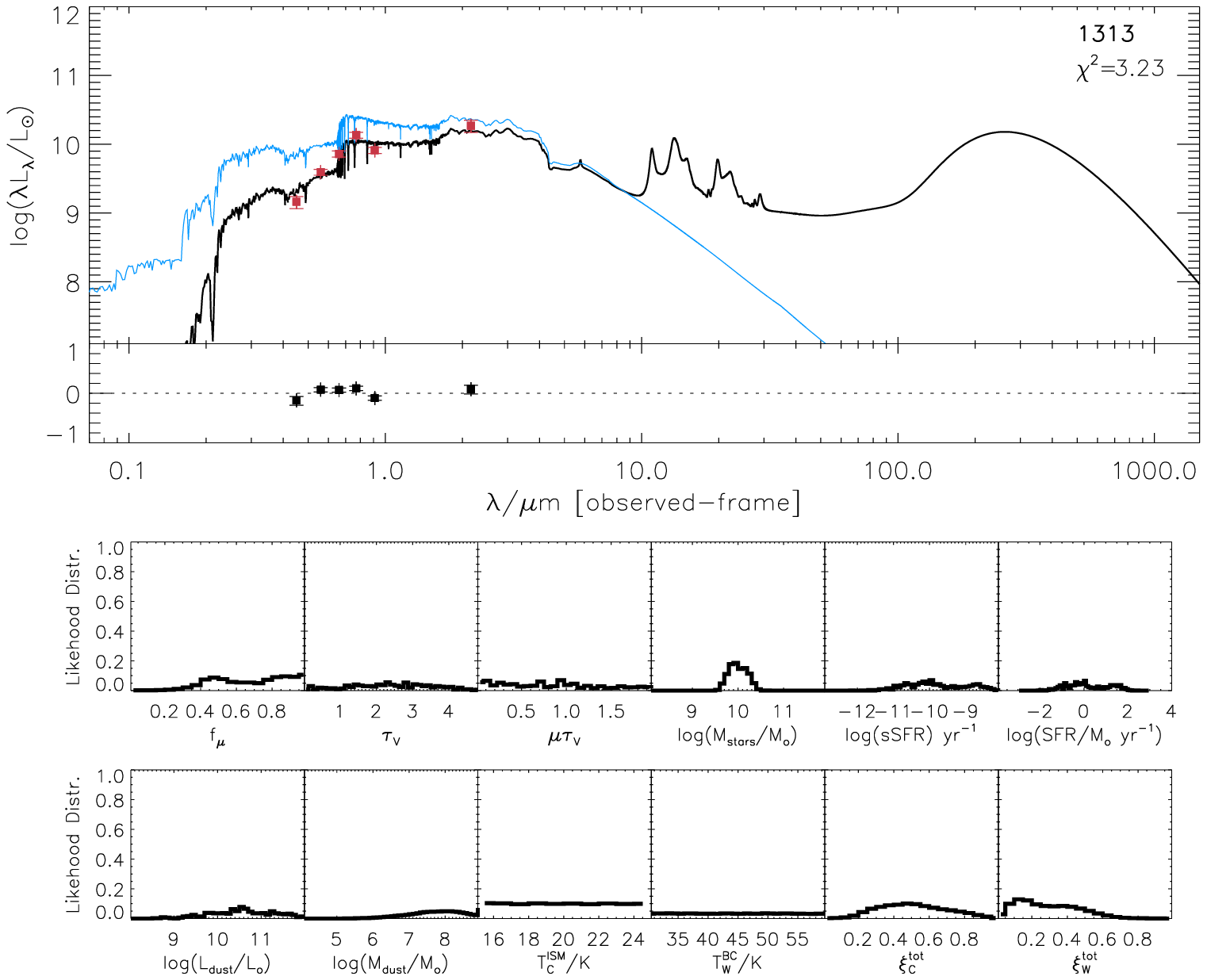}
	\vspace{0.5cm}
	\includegraphics[width=0.95\columnwidth]{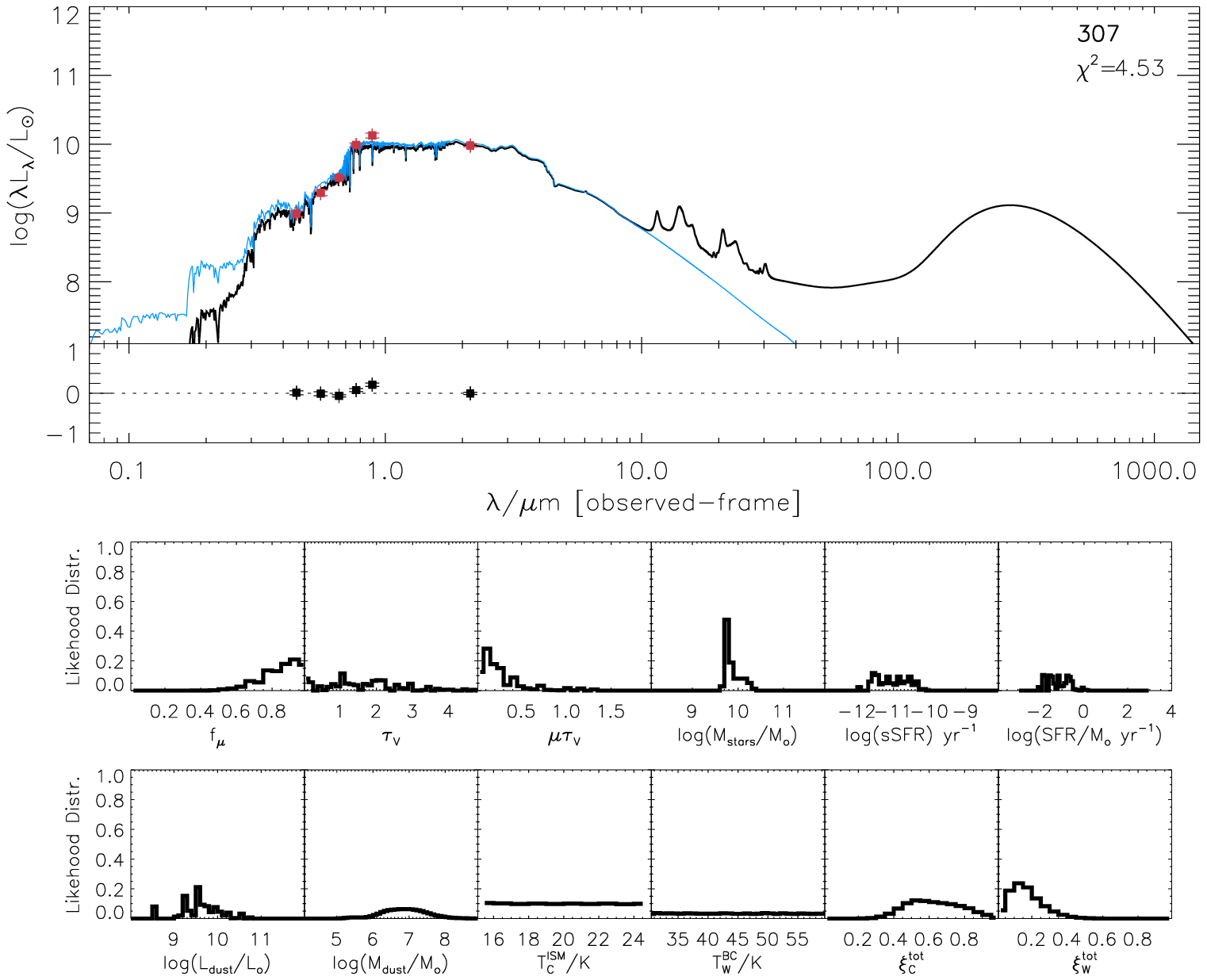}
	\includegraphics[width=0.95\columnwidth]{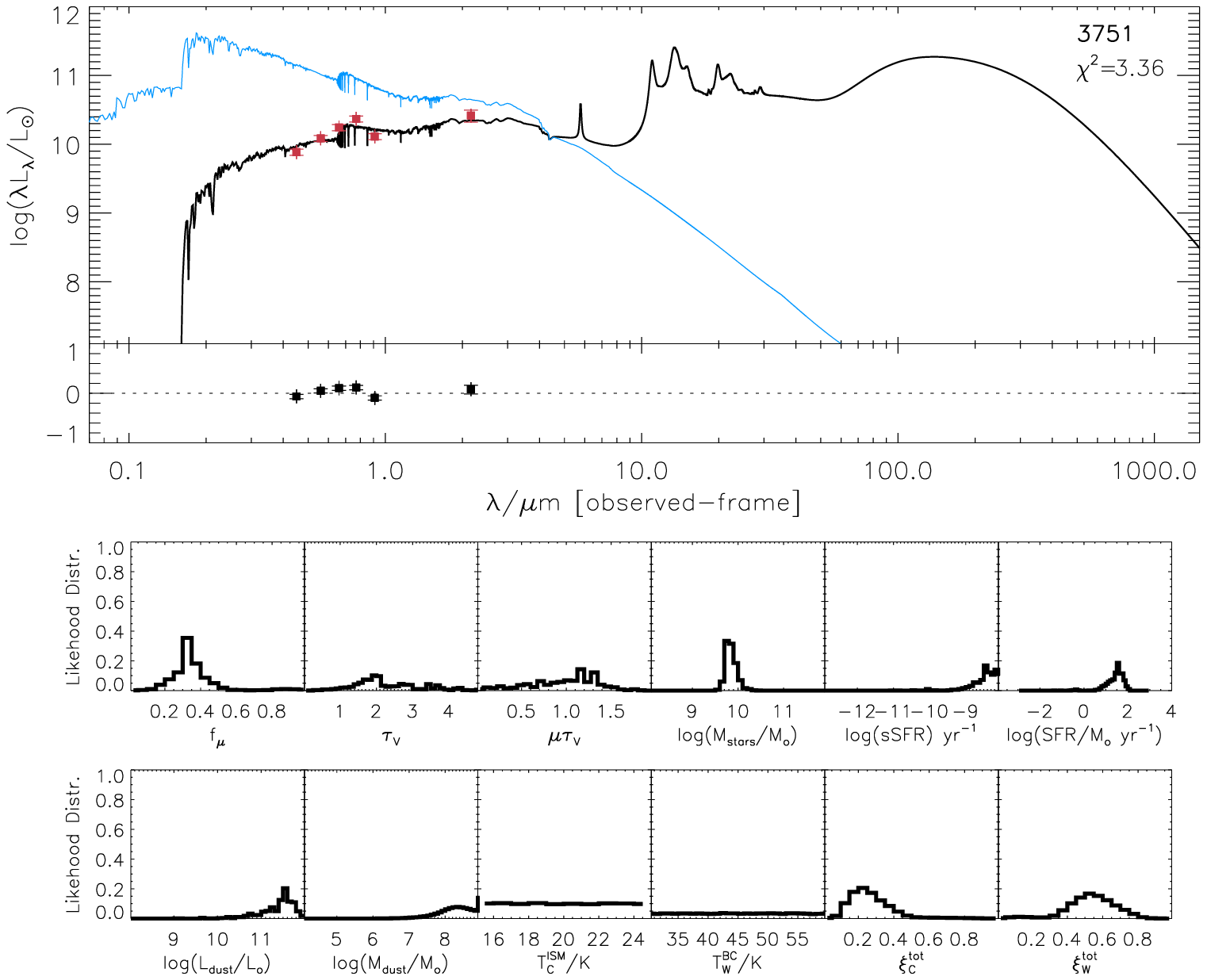}
    \caption{SED fits for high redshift galaxies. {Upper panels}:  typical  ``quenched''  galaxies from  MS 1054-03 and LCDCS 0173. {Lower panels}: typical
    SF galaxies from  CLJ0152 and  LCDCS 0173. Best model fits (black lines) to the observed spectral energy distribution (red points) of the galaxies. In each panel,
    the blue solid line is the the unattenuated stellar population spectrum. The minor panels show the likelihood distribution of the output parameters derived from fits to the observed spectral energy distribution.}
    \label{fig:SEDhighz}    
\end{figure*}

%%%%%%%%%%%%%%%%%%%%%%%%%%%%%%%%%%%%%%%%%%%%%%%%%%

% Don't change these lines
\bsp	% typesetting comment
\label{lastpage}
\end{document}